\documentclass[amsmath,amssymb,aps,10pt,prd,letterpaper,nofootinbib,balancelastpage,notitlepage,superscriptaddress,twocolumn,floatfix,preprintnumbers]{revtex4-2}
\usepackage{graphicx}	
\usepackage{ragged2e}	
\usepackage{bm}		    
\usepackage[sort&compress]{natbib}	 	
	\setcitestyle{square,numbers,comma}	
\usepackage[colorlinks=true,urlcolor=blue,linkcolor=black,citecolor=red]{hyperref}	
\newcommand{\tabref}[2][]{Tab{#1}.~\ref{tab:#2}}		
\newcommand{\figref}[2][]{Fig{#1}.~\ref{fig:#2}}		
\newcommand{\secref}[2][]{Sec{#1}.~\ref{sec:#2}}		
\newcommand{\appref}[2][x]{Appendi{#1}~\ref{app:#2}}	
\renewcommand{\eqref}[2][]{Eq{#1}.~(\ref{eq:#2})}		
\newcommand{\eqrefRange}[2]{Eqs.~(\ref{eq:#1})--(\ref{eq:#2})}		
\newcommand{\citeR}[2][]{Ref{#1}.~\cite{#2}}			
\newcommand{\paragraphdash}[1]{\indent\emph{#1}---\ignorespaces} 
\newcommand{\orcid}[1]{\href{https://orcid.org/#1}{\,\includegraphics[width=8px]{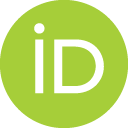}}}
\newcommand{\lb}{\ensuremath{\left}}					
\newcommand{\rb}{\ensuremath{\right}}					
\newcommand{\nl}{\nonumber \\ & \quad }					
\newcommand{\LL}{\mathcal{L}}
\DeclareMathOperator{\IM}{\text{Im}}

\newcommand{\phihat}{\bm{\hat{\phi}}}
\newcommand{\thetahat}{\bm{\hat{\theta}}}
\newcommand{\rhat}{\bm{\hat{r}}}
\newcommand{\Hz}{\,\textrm{Hz}}

\begin{document}

\title{Earth as a transducer for axion dark-matter detection}
\date{\today}
\author{Ariel Arza\orcid{0000-0002-2254-7408}}
\email{ariel.arza@gmail.com}
\affiliation{Institute for Theoretical and Mathematical Physics (ITMP), Lomonosov Moscow State University, 119991 Moscow, Russia}
\author{Michael A.~Fedderke\orcid{0000-0002-1319-1622}}
\email{mfedderke@jhu.edu}
\affiliation{Department of Physics and Astronomy, The Johns Hopkins University, Baltimore, MD 21218, USA}
\author{Peter W.~Graham\orcid{0000-0002-1600-1601}}
\email{pwgraham@stanford.edu}
\affiliation{Stanford Institute for Theoretical Physics, Department of Physics, Stanford University, Stanford, CA 94305, USA}
\affiliation{Kavli Institute for Particle Astrophysics \& Cosmology, Stanford University, Stanford, CA 94305, USA}
\author{Derek F.~Jackson Kimball\orcid{0000-0003-2479-6034}}	
\email{derek.jacksonkimball@csueastbay.edu}
\affiliation{Department of Physics, California State University -- East Bay, Hayward, CA 94542, USA}
\author{Saarik Kalia\orcid{0000-0002-7362-6501}\,}	
\email{saarik@stanford.edu}
\affiliation{Stanford Institute for Theoretical Physics, Department of Physics, Stanford University, Stanford, CA 94305, USA}

\begin{abstract}
We demonstrate that ultralight axion dark matter with a coupling to photons induces an oscillating global terrestrial magnetic-field signal in the presence of the background geomagnetic field of the Earth.
This signal is similar in structure to that of dark-photon dark matter that was recently pointed out and searched for in [\hyperlink{cite.Fedderke:2021rys}{Fedderke~\emph{et~al.}~Phys.~Rev.~D~\textbf{104}, 075023 (2021)}] and [\hyperlink{cite.Fedderke:2021iqw}{Fedderke~\emph{et~al.}~Phys.~Rev.~D \textbf{104}, 095032 (2021)}].
It has a global vectorial pattern fixed by the Earth's geomagnetic field, is temporally coherent on long timescales, and has a frequency set by the axion mass $m_a$.
In this work, we both compute the detailed signal pattern, and undertake a search for this signal in magnetometer network data maintained by the SuperMAG Collaboration.
Our analysis identifies no strong evidence for an axion dark-matter signal in the axion mass range $2\times10^{-18}\,\text{eV} \lesssim m_a \lesssim 7\times10^{-17}\,\text{eV}$.
Assuming the axion is all of the dark matter, we place constraints on the axion--photon coupling $g_{a\gamma}$ in the same mass range; at their strongest, for masses $3\times 10^{-17}\,\text{eV} \lesssim m_a \lesssim 4\times 10^{-17}\,\text{eV}$, these constraints are comparable to those obtained by the CAST helioscope.
\end{abstract}

\maketitle
\tableofcontents

\section{Introduction}
\label{sec:introduction}

The identity of the dark matter (DM)~\cite{zwicky1937masses,rubin1970rotation} remains one of the most prominent unsolved puzzles about the Universe. 
Based on its gravitational effects, we know that the dark matter composes about 26\% of the total energy density of the Universe, the rest consisting of 4\% ordinary baryonic matter and 70\% dark energy~\cite{corbelli2000extended,aghanim2020planck,abbott2019dark,gil2018clustering}. 
To date, no non-gravitational interactions of the dark matter with Standard Model (SM) particles have been observed; if these interactions exist, they must thus be feeble. 
While the ongoing and multi-faceted weakly interacting massive particle (WIMP) detection program continues its long-term~\cite{goodman1985detectability} search for ever-weaker couplings of particle dark matter to the SM, there has recently been increased interest in other dark-matter candidates.
In particular, much attention has been devoted to the study of ultralight classical-field bosonic dark matter, which exhibits distinct and highly varied phenomenology.
The most popular candidates of this type are the QCD axions~\cite{Preskill:1982cy,Abbott:1982af,Dine:1982ah}, axionlike particles (ALPs)~\cite{Gra15,co2020predictions}, and dark photons~\cite{Holdom:1986ag,cvetivc1996implications,Nelson:2011sf}. 
In this paper, we focus our attention on axions.

The QCD axion was originally proposed in order to solve the strong CP problem~\cite{Peccei:1977hh,Weinberg:1977ma,Wilczek:1977pj}. 
Its mass and couplings with SM particles are defined in terms of a single parameter $F_a$, the axion decay constant. 
On the other hand, many extensions to the SM, as well as generic string compactifications, imply the existence of light pseudoscalars with properties similar to the QCD axion~\cite{Svrcek:2006yi,Arvanitaki:2009fg}, but with the crucial difference that their mass and their SM couplings are independent parameters. 
These are usually called axionlike particles (ALPs). 
Both QCD axions and ALPs are produced in the early Universe by non-thermal mechanisms (e.g., misalignment, decays of topological defects, and others~\cite{Preskill:1982cy,Abbott:1982af,Dine:1982ah,Arias:2012az,Co:2017mop,graham2018stochastic,Co:2018mho,Co:2018phi,Co:2019jts,Co:2020dya,Arvanitaki:2019rax,Buschmann:2019icd,Buschmann:2021sdq,Takahashi2018,Takahashi2019}) in sufficient abundance that they can potentially constitute all of the dark-matter energy density (for QCD axions this is true for $F_a$ all the way up to $\sim$ the Planck scale~\cite{graham2018stochastic,Takahashi2018}). 
As these non-thermal production mechanisms create non-relativistic particles, both QCD axions and ALPs are excellent dark-matter candidates~\cite{Preskill:1982cy,Abbott:1982af,Dine:1982ah,Arias:2012az}. 
In this work, we restrict our attention to ALPs, which we will from now on simply refer to as `axions'.

Axions have been mainly searched for via their contribution to electromagnetic signals.
A number of such experiments are based on axion--photon conversion in strong magnetic fields~\cite{Sikivie:1983ip}. These experiments include haloscopes aiming to detect axions from the local axion dark-matter energy density~(see, e.g., \citeR[s]{asztalos2010squid,braine2020extended, Caldwell:2016dcw,zhong2018results,HAYSTAC:2020kwv, Salemi:2021gck,gramolin2021search}), helioscopes to measure a relativistic axion flux coming from the Sun~(see, e.g., \citeR[s]{zioutas2005first,Anastassopoulos:2017ftl,armengaud2019physics}), and light-shining-through-walls (LSW) experiments designed to both produce and detect axions in the laboratory~(see, e.g, \citeR{ehret2010new}). 
It should also be noted that there are a number of new experimental searches that employ the axion--gluon and axion--fermion couplings~\cite{Graham:2013gfa,budker2014proposal,abel2017search,terrano2019constraints,wu2019search,garcon2019constraints,roussy2021experimental,jiang2021search,Ayb21CASPErE}. 
For reviews on axion searches see, e.g., \citeR[s]{Ber05,Fen10,graham2015experimental,safronova2018search,Irastorza:2018dyq,Sikivie:2020zpn}.
 
The magnetic fields involved in many of these searches are produced in the laboratory by ferromagnets or solenoids carrying a strong electric current.
Another possibility is to use instead the natural geomagnetic field of the Earth.%
\footnote{\label{ftnt:previousStudy}%
    One interesting study that exploited the geomagnetic field is \citeR{Davoudiasl:2005nh}, wherein the authors proposed a search for an x-ray signal, on the night-side of the Earth, arising from solar axions converting to photons in the Earth's magnetosphere.
    } %
An important aspect of this field is that its spatial extent is much larger than the length scales that can be achieved in laboratory experiments. 
Axion--photon conversion depends both on the magnitude of the external magnetic field and on the spatial extent over which the interaction takes place. 
Having axion dark matter interacting with the Earth's magnetic field over a length scale of order the Earth's radius thus has the potential to boost axion--photon conversion to a level competitive with experiments that use stronger magnetic fields in a smaller, laboratory-scale volume. 

In this paper, we exploit this observation and point out a novel signal of axion dark matter.
The signal is an oscillating magnetic field generated near the surface of the Earth, and applies for axion masses $m_a \lesssim 10^{-14}\,\text{eV}$.
It is very similar to the one recently pointed out in \citeR[s]{Fedderke:2021rys,Fedderke:2021iqw} for kinetically mixed dark-photon dark matter, which in turn has the same conceptual origin as the signal in the DM Radio experiment~\cite{Chaudhuri:2014dla}.

Consider the axion--photon coupling term in the Lagrangian, in the presence of a static background magnetic field $\bm{B}_0$: $\mathcal{L} \supset + g_{a\gamma} a \bm{E} \cdot\bm{B} \sim g_{a\gamma} (\partial_t a) \bm{B}_0 \cdot \bm{A} \sim -i m_a g_{a\gamma} a \bm{B}_0 \cdot \bm{A}$, where we used $\bm{E} \sim -\partial_t \bm{A}$ (ignoring the EM scalar potential) integrated by parts dropping the boundary term, and used $\partial_t a \sim -i m_a a$ as relevant for a non-relativistic axion.
This term has the same mathematical structure as the mass-mixing term that appears in a Lagrangian describing a dark photon that is kinetically coupled to the SM photon, when expressed in the interaction basis~\cite{Fedderke:2021rys}: $\mathcal{L} \supset \varepsilon m_{A'}^2 A \cdot A' \sim - \varepsilon m_{A'}^2 \bm{A}' \cdot \bm{A} $ (again ignoring the scalar potential), where $m_{A'}$ is the dark-photon mass, and $\varepsilon$ is the kinetic mixing parameter.
The axion coupling must thus source similar physical effects to the dark-photon coupling.
In particular, the well-known effect of the mass-mixing term in the dark-photon case is to create a small mixing of the relevant sterile dark-matter field with the SM electric field in such a way as to drive free-charge motion.
If one considers a shielding conducting box, this results in charges being driven in the conducting walls of the shielding box, and this charge motion will in turn generate real observable electromagnetic fields: in particular, the shield forces the dominant electromagnetic field generated inside the box to be a magnetic field (assuming that the box is of a spatial extent smaller than the dark-photon Compton wavelength).
In exactly the same way, if that box is permeated by a static background magnetic field and the dark matter is instead an oscillating axion field, an oscillating magnetic field is again generated inside the box.
An approximate translation from the dark-photon case to the axion case is simple: $\varepsilon m_{A'}^2 \bm{A}' \rightarrow i g_{a\gamma} m_a a \bm{B}_0$, or, if we assume in each case that the relevant DM candidate is all of the DM, $\varepsilon m_{A'} \bm{\hat{A}}' \rightarrow i g_{a\gamma} B_0 \bm{\hat{B}}_0$.
In particular, the role of the dark-photon polarization state is replaced by the background magnetic-field direction for the axion case, and the axion-induced signal amplitude is obtained from the dark-photon-induced signal amplitude by the replacement $\varepsilon m_{A'}\rightarrow g_{a\gamma} B_0$.

As for the dark-photon case considered in \citeR{Fedderke:2021rys}, our axion signal however does not arise from considering a human-engineered shielding box (as employed, e.g., in other axion searches, such as resonant cavity experiments~\cite{ADMX:2020hay,CAPP:2020utb,HAYSTAC:2020kwv,Alesini:2020vny,McAllister:2017lkb,CAST:2020rlf} and LC circuits~\cite{Sikivie:2013laa,Chaudhuri:2014dla, Salemi:2021gck}); instead, in our mass range of interest, Nature provides us with a ready-made shield.
The near-Earth environment itself can be modeled as a conducting spherical cavity with a vacuum gap~\cite{Fedderke:2021rys}: the lower layer of the atmosphere is a poor conductor sandwiched on one side by the conductive innermost layers of the Earth, and on the other side by the conductive ionosphere and/or interplanetary medium. 
The conductive inner-Earth, ionosphere, and interplanetary medium are thick enough to damp the electromagnetic active mode, at least in our mass range of interest. 
In the case of dark-photon dark matter, this is sufficient to give rise to a dark-matter induced magnetic field at the surface of the Earth~\cite{Fedderke:2021rys}.
Provided that $m_{A'} \ll 1/R$, where $R \sim (3\times 10^{-14}\,\text{eV})^{-1}$ is the radius of the Earth, the electric field was suppressed by a factor $(m_{A'} R)^2$ because of the large natural shield. 
In this paper, we show that because the lower atmospheric vacuum gap is also permeated by the geomagnetic field of the Earth $\bm{B}_{0}$, a similar axion-induced magnetic field is generated if the dark matter is instead composed of axions.
As in the dark-photon case, the accompanying electric field is suppressed by $(m_a R)^{2}\ll 1$, and the magnetic field oscillates at an angular frequency equal to the axion mass $m_a$; it also has a predictable vectorial pattern over the whole surface of the Earth (albeit one that differs from the cognate pattern for the case of dark-photon dark matter), and it inherits the temporal phase-coherence properties of the axion field.

In addition to demonstrating the existence of this novel signal of axion dark matter, we propose to search for it in a way that is conceptually identical to the approach described in \citeR[s]{Fedderke:2021rys,Fedderke:2021iqw}, which advanced and applied this technique for dark-photon dark-matter searches: expose a geographically dispersed network of sensitive magnetometers to the ambient magnetic environment at the surface of the Earth, and record the magnetic field as a function of time over long time periods.
Using distributed networks of sensors%
\footnote{\label{ftnt:GNOME}%
    We note that while there is a geographically distributed array of atomic magnetometers (the Global Network of Optical Magnetometers for Exotic physics searches, GNOME~\cite{Pus13,afach2018characterization,afach2021search}) specifically designed to search for evidence of beyond-the-Standard-Model physics (such as couplings of axion dark-matter fields to nuclear spins~\cite{Pospelov:2012mt}), the GNOME magnetometers are enclosed in meter-scale, multi-layer magnetic shields that effectively cancel the signatures~\cite{Kim16} searched for in this work and that discussed in \citeR[s]{Fedderke:2021rys,Fedderke:2021iqw}.
} %
in this fashion carries many advantages (see, e.g., \citeR[s]{Smiga:2021acs,Chen:2021bdr} for some recent discussions).
This search can make use of the same public database maintained by the SuperMAG Collaboration~\cite{Gjerloev:2009wsd,Gjerloev:2012sdg} that was previously analyzed in \citeR[s]{Fedderke:2021rys,Fedderke:2021iqw}.
SuperMAG collates data from hundreds of unshielded three-axis magnetometers that are widely dispersed over the surface of the Earth and that have been measuring geomagnetic activity since the early 1970s with a time resolution (for the relevant dataset) of one minute.

Because this dataset has previously been analyzed for the dark-photon signal, we can easily motivate why the cognate axion search is interesting.
Ignoring for the purposes of this argument that the Earth's magnetic field takes a non-trivial spatial pattern, we can employ the rough parametric signal-amplitude mapping $\varepsilon m_{A'} \rightarrow g_{a\gamma} B_{0}$ on the existing dark-photon limits set in \citeR{Fedderke:2021rys,Fedderke:2021iqw}.
For instance, for a dark-photon mass of $m_{A'}\sim 4\times10^{-17}\,\text{eV}$, the corresponding smoothed 95\%-credible upper limit on $\varepsilon$ was found in \citeR{Fedderke:2021rys,Fedderke:2021iqw} to be $\varepsilon\sim1.4\times10^{-5}$. 
The Earth's geomagnetic field ranges from $B_{0}\sim25$--$65\,\mu\text{T}$ across its surface,%
\footnote{\label{ftnt:teslaConversion}%
    Recall that $1\,\text{T} \approx 195.4\,\text{eV}^2$.
    } %
so the rough parametric mapping indicates that a bound in the range\linebreak 
$g_{a\gamma} \sim (4.4$--$11)\times10^{-11}\,\text{GeV}^{-1}$ is potentially achievable for $m_a\sim4\times10^{-17}\,\text{eV}$.
At this mass, this estimate brackets the existing low-mass CAST helioscope 95\%-confidence bound on the axion--photon coupling: $g_{a \gamma}\lesssim 6.6\times 10^{-11}\,\text{GeV}^{-1}$~\cite{Anastassopoulos:2017ftl}.
This indicates that it is worthwhile to undertake this analysis carefully, accounting fully for the differing spatial patterns of the dark-photon and axion signals.

Proceeding with a careful analysis of the SuperMAG data, we find no robust evidence for a statistically significant axion-induced oscillating magnetic-field signals.
Our search in the axion mass range $2\times10^{-18}\,\text{eV}\lesssim m_a\lesssim 7\times10^{-17}\,\text{eV}$ initially identifies some 27 na\"ive signal candidates that appear globally significant at the 95\%-confidence level on the basis of our analysis pipeline.
However, further robustness checks performed on these candidates cleanly eliminate the majority of them. 
The only candidates that are not eliminated are either in some tension with some subset of the robustness checks, or have relatively low global significance that would be insufficient to robustly claim anything more than some tension with the background-only model.
Because we find no robust evidence for an axion signal, we proceed to set limits: following a Bayesian analysis procedure that accounts for stochastic fluctuations of the amplitude of the axion dark matter~\cite{Centers:2019dyn,Lisanti:2021vij}, we derive a posterior on the axion--photon coupling and place 95\%-credible upper limits on $g_{a\gamma}$ in the same mass range as for the signal search.
Assuming that the axion is all of the dark matter, our limits indeed reach the current CAST bound at their most sensitive: we set the constraint $g_{a\gamma} \lesssim 6.5\times 10^{-11}\,\text{GeV}^{-1}$ for $3\times 10^{-17}\,\text{eV}\lesssim m_a \lesssim 4\times 10^{-17}\,\text{eV}$ (the limits weaken outside this range).
Nevertheless, our limits have distinct systematics as compared to CAST, and future improvements using other existing archival datasets as well as via dedicated searches with new experiments are possible.

The rest of this paper is structured as follows:
in \secref{signal}, we derive the axion dark-matter induced magnetic-field signal, beginning with a derivation of the axion effective current in \secref{effectivecurrent}, then discussing the IGRF-13 geomagnetic-field model in \secref{igrfmodel}, giving a quick signal derivation argument in \secref{derivation} with many details deferred to the appendices, and then comparing it to the cognate dark-photon signal~\cite{Fedderke:2021rys} in \secref{comparison}.
In \secref{analysis}, we summarize our signal search at a high level, again deferring details to the appendices, and present a set of axion--photon coupling exclusion bounds in \figref{limit}.
We conclude in \secref{conclusion}.
There are a number of appendices that expand on the main text with further detail:
\appref{vectorSphericalHarmonics} gives our conventions for the vector spherical harmonics and a number of relevant identities that we utilize in this work.
In \appref{detailedcalculation}, we present a thorough and detailed derivation of the signal under two different sets of assumptions regarding the modeling of the near-Earth conductivity environment; this reinforces the quicker derivation given in the main text.
The details of our analysis are presented in \appref{analysisdetails}; we mirror the discussion in \citeR{Fedderke:2021iqw} and give a minute accounting of all relevant differences.
Included in \appref{reevaluation} is a detailed investigation of some anomalies (`na\"ive signal candidates') that we identified in the data, but which we do not consider to be strong, robust signals of axion dark matter for reasons also discussed in that appendix.

\section{Signal}
\label{sec:signal}

In this section, we describe the observable magnetic-field signal sourced by axion dark matter at the surface of the Earth.
The signal described in this work is analogous to the one described in \citeR{Fedderke:2021rys}, but with axion dark matter replacing the role of dark-photon dark matter.

It can be shown that the Earth itself acts as an effective conducting shield for electromagnetic waves of frequencies $10^{-21}\,\text{eV}\lesssim\omega\lesssim3\times10^{-14}\,\text{eV}$ (see Sec.~II\,B of \citeR{Fedderke:2021rys}).
At these frequencies, the lower atmosphere just above the surface of the Earth, however, has negligible damping effects on electromagnetic waves~\cite{Fedderke:2021rys}.
Even further from the surface of the Earth, either the ionosphere or interplanetary medium act as an effective shield in this frequency range~\cite{Fedderke:2021rys} owing to their large conductivity or plasma frequency, respectively.%
\footnote{\label{ftnt:conductivity}%
    In roughly the range $(\text{few})\times10^{-16}\,\text{eV}\lesssim\omega\lesssim3\times10^{-14}\,\text{eV}$, the ionosphere acts as an effective shield.
    In the range $10^{-21}\,\text{eV}\lesssim\omega\lesssim(\text{few})\times10^{-16}\,\text{eV}$, the effects of the ionosphere become more complicated, and so it is the interplanetary medium that can be considered the outer shield. 
    See Sec.~II\,B of \citeR{Fedderke:2021rys} for a detailed discussion of the electromagnetic behavior of the ionosphere.
} %
The near-Earth environment can thus be treated as an inner conducting sphere (the Earth) and a surrounding conducting shield (the ionosphere/interplanetary medium), separated by a vacuum region (the lower atmosphere).

In \citeR{Fedderke:2021rys}, the effect of kinetically mixed dark-photon dark matter in this environment was parametrized by an effective background current, oriented in the direction of the local dark-photon field.
It was shown that the effect of this current would be to generate an oscillating magnetic-field signal at the surface of the Earth which exhibits a particular global spatial pattern, given by $\bm\Phi_{\ell m}$ vector spherical harmonics (VSH) [see \appref{vectorSphericalHarmonics} for VSH conventions].
Importantly, the leading $\bm\Phi_{\ell m}$ contributions did not depend on the details of the conducting boundaries.

The derivation in this work of the signal of axion dark matter in this near-Earth conductivity environment proceeds similarly to the derivation for dark-photon dark matter in \citeR{Fedderke:2021rys}, but with some important differences.
Specifically, we can also employ an effective current approach for electromagnetically coupled axion dark matter.
One crucial difference between the axion and dark-photon cases however is that the axion requires the Earth's static magnetic field in order to convert into an observable electromagnetic signal.
The axion-induced effective current will thus not only depend on the local axion field value but also on the local geomagnetic field.
In particular, the effective current will inherit its direction from the Earth's quasi-static magnetic field.
Once translated into the language of an effective current, the axion calculation proceeds similarly to the dark-photon calculation.
We thus defer a detailed calculation of the axion dark-matter signal to \appref{detailedcalculation}, and instead in this section present a simpler argument (which could also apply to the dark-photon case).

We begin this section with a description of the effective current for axion dark matter in the presence of a magnetic field.
Next, we describe the International Geomagnetic Reference Field (IGRF) model for the Earth's static magnetic field.
Then we present a simple derivation for the $\bm\Phi_{\ell m}$ component of the axion dark-matter signal.
Finally, we conclude by comparing the properties of this axion signal with the previously described dark-photon signal.

\subsection{Effective current}
\label{sec:effectivecurrent}

In this work, we consider an axion $a$ coupled to electromagnetism with strength $g_{a\gamma}$, described by the Lagrangian
\begin{align}
    \LL\supset\frac12(\partial_\mu a)^2-\frac12m_a^2a^2-\frac14F_{\mu\nu}F^{\mu\nu}+\frac{1}{4}g_{a\gamma}a F_{\mu\nu}\widetilde{F}^{\mu\nu}.
\end{align}
In the presence of such an axion, Maxwell's equations are modified as~\cite{Sikivie:1983ip,Wilczek:1987edt}
\begin{align}
    \label{eq:Gauss_Law}
    \nabla\cdot\bm E&=g_{a\gamma}\nabla a\cdot\bm B,\\
    \nabla\cdot\bm B&=0,\\
    \nabla\times \bm E+\partial_t\bm B&=0,\label{eq:Faradays_Law}\\
    \nabla\times\bm B-\partial_t\bm E&=-g_{a\gamma}\lb[ (\partial_t a)\bm B+\nabla a\times\bm E \rb].
    \label{eq:Amperes_Law}
\end{align}
If the axion $a$ comprises the dark matter, then it is non-relativistic and so $\partial_ta\sim-im_aa$ and $|\nabla a|\sim m_av_\textsc{dm}a$, where $v_\textsc{dm}\sim10^{-3}$.
Therefore, all $\nabla a$ terms will be parametrically suppressed compared to the $\partial_t a$ terms, and
the leading effect of the axion dark matter will thus come from the $-g_{a\gamma}(\partial_t a)\bm B$ term in \eqref{Amperes_Law}.
We can see from the form of \eqref{Amperes_Law} that this term behaves similarly to a current in the usual Amp\`ere-Maxwell law,
\begin{align}
    \nabla\times\bm B-\partial_t\bm E=\bm J_\text{eff}.
    \label{eq:Amperes_eff}
\end{align}
In the presence of a static background magnetic field $\bm B_0$, this effective current is given by
\begin{align}
    \bm J_\text{eff}=ig_{a\gamma}a m_a\bm B_0.
    \label{eq:eff_current}
\end{align}
Note that in the case of dark-photon dark matter, the direction of $\bm J_\text{eff}$ is set by the direction of the dark photon polarization~\cite{Fedderke:2021rys} and is thus spatially uniform. By contrast, in the axion case, the direction of $\bm J_\text{eff}$ is not determined by any property of the axion, but rather by the direction of the static magnetic field $\bm B_0$.
For the derivation of our signal, $\bm B_0$ will be the Earth's DC magnetic field.
Thus, the effective current in the axion case will not be uniform in space, but will have an (approximately) dipolar angular dependence and decay with radial distance from the Earth's center.

\subsection{IGRF model}
\label{sec:igrfmodel}

In this work, we employ the IGRF-13 model~\cite{IGRF} of the Earth's magnetic field, which provides coefficients for a multipole expansion of the field.
As the geomagnetic field drifts slowly over time, the IGRF model provides coefficients for the field at five-year intervals and specifies an interpolation procedure on these coefficients to obtain the field at intermediate times.
The most recent generation, IGRF-13, provides values dating from 1900 up to 2020.
The IGRF model parametrizes the geomagnetic field in terms of a scalar potential%
\footnote{\label{ftnt:magnetic_potential}%
    Recall that in the magnetoquasistatic limit and in the absence of free currents, the Amp\`ere-Maxwell law becomes $\nabla\times\bm B=0$.
    We can therefore define a scalar potential $V$ so that $\bm B=-\nabla V$.
    See Sec.~5.9.B of \citeR{Jackson} for a more detailed discussion.
} %
$\bm B_0=-\nabla V_0$, which is then expanded as
\begin{align}
    V_0&=\sum_{\ell=1}^\infty\sum_{m=0}^\ell\frac{R^{\ell+2}}{r^{\ell+1}}\left(g_{\ell m}\cos m\phi+h_{\ell m}\sin m\phi\right)P_\ell^m(\cos\theta)\\
    &=\sum_{\ell=1}^\infty\sum_{m=-\ell}^\ell(-1)^m\sqrt{\frac{4\pi(2-\delta_0^m)}{2\ell+1}}\frac{R^{\ell+2}}{r^{\ell+1}}\frac{g_{\ell m}-ih_{\ell m}}2 \nl\qquad\qquad\quad \times  Y_\ell^m(\theta,\phi),
\end{align}
where $R=6371.2\,$km is the exact reference value for the Earth's radius that is used in the specification of the IGRF model~\cite{IGRF}, $\theta$ and $\phi$ are geographic co-latitude and longitude,%
\footnote{\label{ftnt:corotating-frame}%
    Note that the coordinates $\theta$ and $\phi$ co-rotate with the Earth; that is, they describe a rotating frame with coordinates fixed to the Earth, not the inertial frame with coordinates fixed to the average positions of distant stars.
    Henceforth, all VSH and spherical harmonics will implicitly use coordinates in this co-rotating frame as well.
    As the rotational speed at the surface of the Earth is non-relativistic, there are no relativistic field-mixing effects which need to be taken into account when switching between these frames.
    Thus, it remains consistent to apply Maxwell's equations as in \eqrefRange{Gauss_Law}{Amperes_Law} even in the co-rotating frame.
    Specifically, because \emph{both} the Earth's magnetic field and the observation points for the magnetic observatories co-rotate with the Earth, this is the natural co-ordinate frame to use.
} %
$P_\ell^m$ are the \emph{Schmidt-normalized} associated Legendre polynomials,%
\footnote{\label{ftnt:Schmidt-norm}%
    These are given for $m\geq0$ by~\cite{Winch2005}
    \begin{align}
        P_\ell^m(x)=\sqrt{(2-\delta_0^m)\frac{(\ell-m)!}{(\ell+m)!}}\left(1-x^2\right)^{m/2}\frac{d^m}{dx^m}P_\ell(x),
    \end{align}
    where $P_\ell$ are the Legendre polynomials.
    Note that the Schmidt-normalized $P_{\ell}^{m}$ used in the IGRF specification differ from those defined at Eq.~(3.49) in \citeR{Jackson} in several ways; however, our scalar $Y_{\ell}^{m}$ are normalized to agree with those of \citeR{Jackson}.
} %
and $Y_\ell^m$ are the scalar spherical harmonics.
The IGRF model provides the `Gauss coefficients' $g_{\ell m}$ and $h_{\ell m}$ for $\ell\geq m\geq0$ at five-year intervals (see Tab.~2 of \citeR{IGRF}), and their values at intermediate times are to be calculated by linear interpolation.
Here we adopt the conventions $g_{\ell,-m}=(-1)^mg_{\ell m}$ and $h_{\ell,-m}=(-1)^{m+1}h_{\ell m}$ to extend the coefficients to negative $m$. 
From $\bm B_0=-\nabla V_0$, we can then write a multipole expansion for the geomagnetic field in terms of the VSH as
\begin{align}
    \bm B_0=\sum_{\ell,m}C_{\ell m} \left(\frac Rr\right)^{\ell+2}\big[(\ell+1)\bm Y_{\ell m}-\bm\Psi_{\ell m}\big],
    \label{eq:Earth_field}
\end{align}
where $C_{\ell m}$ are related to the Gauss coefficients by
\begin{align}
    C_{\ell m}=(-1)^m\sqrt{\frac{4\pi(2-\delta_0^m)}{2\ell+1}}\frac{g_{\ell m}-ih_{\ell m}}2.
    \label{eq:clm}
\end{align}
Note that our phase conventions for $g_{\ell m}$ and $h_{\ell m}$ ensure that $C_{\ell,-m}=(-1)^mC_{\ell m}^*$, in analogy to the VSH phase conventions \eqrefRange{Yminus}{Phiminus}.
As the Earth's magnetic field is approximately dipolar, with the dipole axis oriented relatively close to the Earth's rotational axis, the largest of these coefficients will be $C_{10}$; however, subsequent terms can provide corrections of $\mathcal O(10\%)$.
The IGRF-13 model provides values up to $\ell=13$ for the most recent coefficients.
In our analysis (summarized in \secref{analysis} and detailed in \appref{analysisdetails}), we find it sufficient to utilize the Gauss coefficients up to $\ell=4$.
We have verified explicitly that the addition of higher-$\ell$ modes has no significant impact on our results.

\subsection{Signal derivation}
\label{sec:derivation}

Here we give a simple derivation for the leading order $\bm\Phi_{\ell m}$ contribution to the magnetic-field signal of axion dark matter at the Earth's surface (see \appref{detailedcalculation} for a more detailed calculation).
This derivation relies only on the effective current approach, and so a similar derivation can also be applied to the dark-photon case computed in \citeR{Fedderke:2021rys}.
As in \citeR{Fedderke:2021rys}, we model the near-Earth environment as a perfectly conducting sphere of radius $R$ (the Earth), surrounded by some vacuum region (the lower atmosphere), which is further surrounded by some perfectly conducting boundary.
Here we do not assume any particular shape for the outer boundary, only that it has a longest length scale $L\ll m_a^{-1}$.
For this reason, our model allows for the outer boundary to be either the ionosphere, which is approximately spherical and located only $\sim100\,\text{km}$ from the Earth's surface, or the magnetopause (beyond which lies the interplanetary medium), which is highly aspherical and extends to $\sim200R$ from the Earth's surface in the `downwind' direction of the solar wind.%
\footnote{\label{ftnt:model_breakdown}%
    We note that for $m_a\sim(\text{few})\times10^{-16}\,\text{eV}$, which is the boundary between the frequency ranges mentioned in footnote \ref{ftnt:conductivity}, the damping effects of the ionosphere are uncertain, yet the furthest point of the magnetopause may not be a sub-wavelength distance from the Earth's surface; i.e., $200R\sim m_a^{-1}$.
    The validity of our assumptions are therefore questionable around $m_a\sim(\text{few})\times10^{-16}\,\text{eV}$; see also the discussion in Sec.~II\,B (and in particular Sec.~II\,B\,6) of \citeR{Fedderke:2021rys}.
    This mass lies outside of the range explicitly constrained in our analysis (see \secref{analysis}).
} %
As this longest length scale $L$ is smaller than the de Broglie wavelength%
\footnote{\label{ftnt:dBwavelength}%
    In principle, there are two length scales of the dark matter which could be relevant here: the de Broglie wavelength $\lambda_\text{dB}\sim(m_av_\text{rel})^{-1}$ and the coherence length $\lambda_\text{coh}\sim(m_a\Delta v)^{-1}$, where $v_\text{rel}$ is the mean relative velocity between the dark-matter rest frame and the Earth, and $\Delta v$ is the magnitude of the local DM velocity dispersion.
    In order for \eqref{axion} to remain valid, $L$ must be smaller than \emph{both} of these length scales.
    We note however that for virialized dark matter following a Maxwell--Boltzmann velocity distribution, $v_\text{rel}\sim\Delta v\sim10^{-3}$ in the Earth's rest frame, so that $\lambda_\text{dB}\sim\lambda_\text{coh}$.
} %
of the axion dark matter $\lambda_\text{dB}\sim(m_av_\textsc{DM})^{-1}$, we can take the axion field value to be constant over the entire geometry, and write it as
\begin{align}
    a=a_0e^{-im_a t}.
    \label{eq:axion}
\end{align}
Due to the stochasticity of the axion field~\cite{Foster:2017hbq,Centers:2019dyn}, $a_0$ is not uniquely determined by the DM density, although it is generically of order $|a_0| \sim \sqrt{2\rho_\textsc{dm}}/m_a$ (see discussion at the end of this subsection).

From \eqref[s]{eff_current} and (\ref{eq:Earth_field}), the effective current which this axion dark matter sources is then given by
\begin{align}
    \bm J_\text{eff}&=ig_{a\gamma}a_0m_a\nl
    \; \times\sum_{\ell,m}C_{\ell m}\left(\frac Rr\right)^{\ell+2}\big[(\ell+1)\bm Y_{\ell m}-\bm\Psi_{\ell m}\big]e^{-im_a t}.
    \label{eq:Eartheff_current}
\end{align}

Now we argue that the $\partial_t\bm E$ term in \eqref{Amperes_eff} can be neglected.%
\footnote{\label{ftnt:neglectE}%
    See the end of Sec.~III\,C in \citeR{Fedderke:2021rys} for a similar discussion.
} %
This is because $\bm E$ vanishes both deep within the Earth and within a skin depth of the outer boundary (as they are both good enough conducting shields to effectively damp all electromagnetic waves).
Moreover, as our geometry has longest length scale $L\ll m_a^{-1}$, these two surfaces on which $\bm{E}$ vanishes are separated by a sub-wavelength distance.
We therefore only expect that $\bm E$ can grow quadratically in $m_a L$ between them.
In particular, we expect parametrically $\bm E\sim(g_{a\gamma}a_0)(m_a L)^2C_{\ell m}$.
Comparing to \eqref{Eartheff_current}, we see that $\partial_t\bm E$ is parametrically smaller than $\bm J_\text{eff}$.
Therefore, up to corrections at order $\mathcal O((m_a L)^2)$, it suffices to only consider the first and last terms of \eqref{Amperes_eff}.

Given the form of $\bm J_\text{eff}$ in \eqref{Eartheff_current}, we can apply the VSH curl properties \eqrefRange{Ycurl}{Phicurl} to solve \eqref{Amperes_eff}.
Namely, we find that $\bm B$ must be of the form
\begin{align}
    \bm B&=-i(g_{a\gamma}a_0)(m_a R)\sum_{\ell,m}\frac{C_{\ell m}}\ell\left(\frac Rr\right)^{\ell+1}\bm\Phi_{\ell m}e^{-im_a t}\nl
    \quad+\nabla V+\mathcal O\lb(\lb(m_a L\rb)^2\rb),
    \label{eq:full_magnetic}
\end{align}
where $V$ is some scalar function (so that $\nabla V$ is curl-free).
From the spherical-harmonic gradient relation \eqref{gradient}, we can also note that $\nabla V$ consists entirely of $\bm Y_{\ell m}$ and $\bm\Psi_{\ell m}$ modes.
Therefore, the leading order $\bm\Phi_{\ell m}$ contribution to the magnetic field is precisely given by the first line of \eqref{full_magnetic}.
In particular, at the surface of the Earth~($r=R$), to leading order in $m_a L$, the $\bm\Phi_{\ell m}$ contribution to the magnetic field signal of axion dark matter is%
\footnote{\label{ftnt:TM_mode}%
    Here, we label the $\bm\Phi_{\ell m}$ contribution as `transverse magnetic' (TM).
    See the discussion around \eqrefRange{ETE}{ETM} or Sec.~III\,B of \citeR{Fedderke:2021rys} for an explanation of transverse magnetic and transverse electric (TE) modes.
} %
\begin{align}
    \bm B_\text{TM}=-i(g_{a\gamma}a_0)&(m_a R)\sum_{\ell,m}\frac{C_{\ell m}}\ell\bm\Phi_{\ell m}e^{-im_a t}.
    \label{eq:signal}
\end{align}
Note that due to the VSH orthogonality properties \eqrefRange{Y_orthog}{cross_orthog}, any vectorial function on the sphere can be decomposed into VSH (much like any scalar function on the sphere can be decomposed into scalar spherical harmonics).
Thus, when searching for our signal in global magnetic-field data across the Earth, we can project onto the particular combination of $\bm\Phi_{\ell m}$ modes appearing in \eqref{signal}.
This allows us to neglect the $\bm Y_{\ell m}$ and $\bm\Psi_{\ell m}$ contributions coming from $\nabla V$, which may generically depend on the shape of the outer boundary, and instead focus on the $\bm\Phi_{\ell m}$ contributions which we know to be present regardless of details of the outer boundary.

Finally, we comment on the temporal coherence of our signal in \eqref{signal}.
The monochromatic description of the axion given in \eqref{axion} remains valid only on timescales less than the coherence time $T_\text{coh}\sim2\pi/(m_a v_\textsc{dm}^2)$ of the axion.
For the mass range relevant to our analysis (summarized in \secref{analysis} and detailed in \appref{analysisdetails}), we have $T_\text{coh}\sim 2$--$50\,$yr.
On timescales longer than the coherence time, $a_0$ will vary stochastically in both amplitude and phase.
Therefore, \eqref{signal} only remains valid for times $t\lesssim T_\text{coh}$: the magnetic-field signal's phase offset and amplitude randomize on longer timescales, with the phase offset within each coherence time being uniformly distributed on $[0,2\pi)$, and the amplitude being set by $a_0$ drawn from a distribution~\cite{Centers:2019dyn,Foster:2017hbq} and satisfying $\langle|a_0|^2 \rangle_{\tau} = 2\rho_{\textsc{dm}}/m_{a}^2$, on average over timescales $\tau \gg T_{\text{coh}}$. 
See also the more detailed discussion of this point in the context of dark-photon dark matter in \citeR{Fedderke:2021rys}.

\subsection{Comparison with dark-photon signal}
\label{sec:comparison}

The signal described by \eqref{signal} takes a very similar form to the signal described in \citeR{Fedderke:2021rys}.
In particular, if the Earth's magnetic field is assumed to be exactly dipolar ($C_{\ell m}=0$ for $\ell>1$), then \eqref{signal} takes precisely the same form as the signal from a dark photon expressed in \emph{inertial coordinates}, with the role of the dark-photon polarization in setting the signal orientation replaced by the Earth's magnetic dipole (cf.~Eq.~(38) of \citeR{Fedderke:2021rys}). 
Here we highlight three important differences that allow an axion dark-matter signal to be distinguished from a dark-photon signal.

The first is due to the fact that the geomagnetic field is not exactly dipolar, and so $\ell>1$ modes will contribute to \eqref{signal}, giving it a slightly different angular dependence than a dark-photon signal.
As mentioned before, this correction in angular dependence will be at the level of $\mathcal O(10\%)$.

Secondly, the dark-photon signal \emph{in co-rotating coordinates} receives a shift in frequency by $f_d=(\text{sidereal day})^{-1}$ due to the rotation of the Earth (see Eq.~(42) of \citeR{Fedderke:2021rys}), whereas the axion dark-matter signal does not.
This is because the effective current (and thus the angular dependence of the magnetic-field signal) inherits its direction from the geomagnetic field in the axion case but from the dark-photon field itself in the dark-photon case.
Because the geomagnetic field co-rotates with the Earth, the angular dependence of the axion signal is constant in geographic coordinates (on timescales short enough that the geomagnetic field does not drift significantly).
On the other hand, since the dark-photon direction is fixed in inertial coordinates (on timescales shorter than the coherence time of the dark-photon field), the dark-photon signal precesses in geographic coordinates.
Thus, in frequency space, the axion signal (as measured by magnetometers fixed on the Earth's surface) only appears at $f=f_a\equiv m_a/2\pi$ (assuming $v_{\textsc{dm}}=0$; see footnote 34 in \citeR{Fedderke:2021iqw}), while the dark-photon signal also exhibits sidebands at $f=f_a\pm f_d$.

Finally, the stochastic properties of the axion dark-matter signal could differ from those of the dark-photon signal.
Classical-field dark-matter candidates (both axions and dark photons) are comprised of a sum (really, an integral) over constituent Fourier modes, each of which has a random phase and, for the dark-photon case, a vectorial orientation (which is in general complex).
As a result, classical-field dark-matter exhibits amplitude and overall phase-offset fluctuations from one coherence time to the next; see, e.g., \citeR[s]{Foster:2017hbq,Centers:2019dyn}.
In addition, for the dark-photon case, there can be a fluctuation of the polarization state of the field, but this depends on the assumed underlying structure of the individual Fourier modes' vectorial orientations.
Depending on the formation model and subsequent cosmological evolution of the dark matter, it is an open question (see, e.g., \citeR{Caputo:2021eaa}) whether these individual Fourier modes' vectorial orientations are all the same, or whether they are effectively random.
In the former case, the dark-photon polarization state does not randomize from one coherence time to the next; in the latter case, it does.%
\footnote{\label{eq:ourDPDManalysis}%
    Formally, in our analysis in \citeR[s]{Fedderke:2021iqw,Fedderke:2021rys} we assumed this latter case.
    However, because the network of magnetometers contributing to the SuperMAG dataset has reasonably isotropic directional sensitivity (being a network of $\mathcal{O}(500)$ three-axis magnetometers that are widely distributed on the rotating and orbiting Earth), we expect that even if the dark-photon dark matter behaved according to the former case, our limits in \citeR[s]{Fedderke:2021iqw,Fedderke:2021rys} would be changed by only an $\mathcal{O}(1)$ factor.
} %

Since the direction of the effective current that gives rise to the dark-photon signal is set by the dark-photon polarization state, the polarization-state fluctuation that arises in this latter case results in a fluctuation of the relative phases appearing between the different components of the effective current from one coherence time to the next.
Because the magnetic-field signal is determined by the effective current, this results in an $\mathcal{O}(1)$ change to the global angular dependence of the dark-photon-induced magnetic-field signal from one coherence time to the next in this case. 

By contrast, for axion dark matter, the direction of the effective current is determined by the geomagnetic field, so no fluctuation in the relative phases of the components of the effective current appears from one coherence time to the next.
The direction of the effective current and the global angular dependence of the magnetic-field signal instead drift only as the geomagnetic field drifts on very long timescales (which is independent of the coherence properties of the axion).

Over the 50-year duration of the SuperMAG dataset utilized in our analysis (summarized in \secref{analysis} and detailed in \appref{analysisdetails}), the direction of the geomagnetic field drifts $\mathcal O(5\%)$.
Thus, the change in global angular dependence of the axion dark-matter signal is significantly smaller over the duration of our analysis than the $\mathcal O(1)$ change that the dark-photon signal experiences over timescales of order the coherence time, in the case assumed above.
Moreover, the Earth's background magnetic-field drift is modeled, and we include this effect in our analysis of the axion signal, whereas the possible coherence-time to coherence-time drift in the global angular dependence of the signal in the dark-photon case is inherently stochastic, although accounted for in~\citeR{Fedderke:2021iqw}.

\section{Search for signal in SuperMAG dataset}
\label{sec:analysis}

The axion dark-matter signal described by \eqref{signal} is an oscillating magnetic field at the surface of the Earth of magnitude
\begin{align}
    B \sim 1\,\text{nG} \times \lb( \frac{g_{a\gamma}}{10^{-10}\,\text{GeV}^{-1}} \rb),
\end{align}
assuming the axion constitutes all of the dark matter (we take $\rho_{\textsc{dm}}=0.3\,\text{GeV/cm}^3$).
It is temporally coherent over a long time period; it is also spatially coherent, taking a known pattern across the entire globe.
As such, it could be detected by any global array of unshielded magnetometers taking data over several decades.
The SuperMAG Collaboration~\cite{Gjerloev:2009wsd,Gjerloev:2012sdg,SuperMAGwebsite} maintains a public database of measurements from precisely such an array of magnetometers.
In particular, they report three-axis magnetic field measurements from $\mathcal O(500)$ stations with a one-minute time resolution, with the measurements from some stations dating back to 1970.

In this work, we perform a search of the SuperMAG dataset%
\footnote{\label{ftnt:updated_dataset}%
    In this work, we use a slightly updated dataset as compared to the one used in \citeR{Fedderke:2021iqw}, which includes a few additional stations.
    The temporal duration of the dataset remains the beginning of 1970 through the end of 2019.
    As this updated dataset contains minimal additional data, we do not expect to gain significant sensitivity.
} %
for the signal described by \eqref{signal}, similar to the dark-photon dark-matter search undertaken in \citeR{Fedderke:2021iqw}.
As the analysis proceeds in a similar way to the one in \citeR{Fedderke:2021iqw}, we reserve the details of this work's analysis (and, in particular, how they differ from those of the analysis in \citeR{Fedderke:2021iqw}) to \appref{analysisdetails}.
Instead, in this section, we summarize the results of this work's axion dark-matter search.
We first enumerate some na\"ive candidate signals which we identified but through further robustness checks dismissed.
Having dismissed all such na\"ive signal candidates, we then present an exclusion limit on axion dark-matter parameter space.
Finally, we discuss a higher-resolution dataset also maintained by the SuperMAG collaboration, a future analysis of which could extend the results of this work to new parameter space.

As in \citeR{Fedderke:2021iqw}, we performed a search for our signal at $\mathcal O(10^6)$ discrete frequencies in the frequency range $6\times10^{-4}\,\text{Hz}\lesssim f_a\lesssim 2\times10^{-2}\,\text{Hz}$, corresponding to the mass range $2\times10^{-18}\,\text{eV}\lesssim m_a\lesssim 7\times10^{-17}\,\text{eV}$.
For each frequency, we constructed analysis variables (see \appref{timeseries}) and wrote down a likelihood function for the photon-axion coupling $g_{a\gamma}$, given the observed values of these variables (see \appref{bayesian}).
Using this likelihood, we determined whether the analysis variables at each frequency were consistent with the lack of a signal $g_{a\gamma}=0$.
We declared any frequency at which the data were inconsistent with $g_{a\gamma}=0$ at 95\% confidence global significance to be a `na\"ive signal candidate' for axion dark matter.

\begin{figure*}[t]
\includegraphics[width=\textwidth]{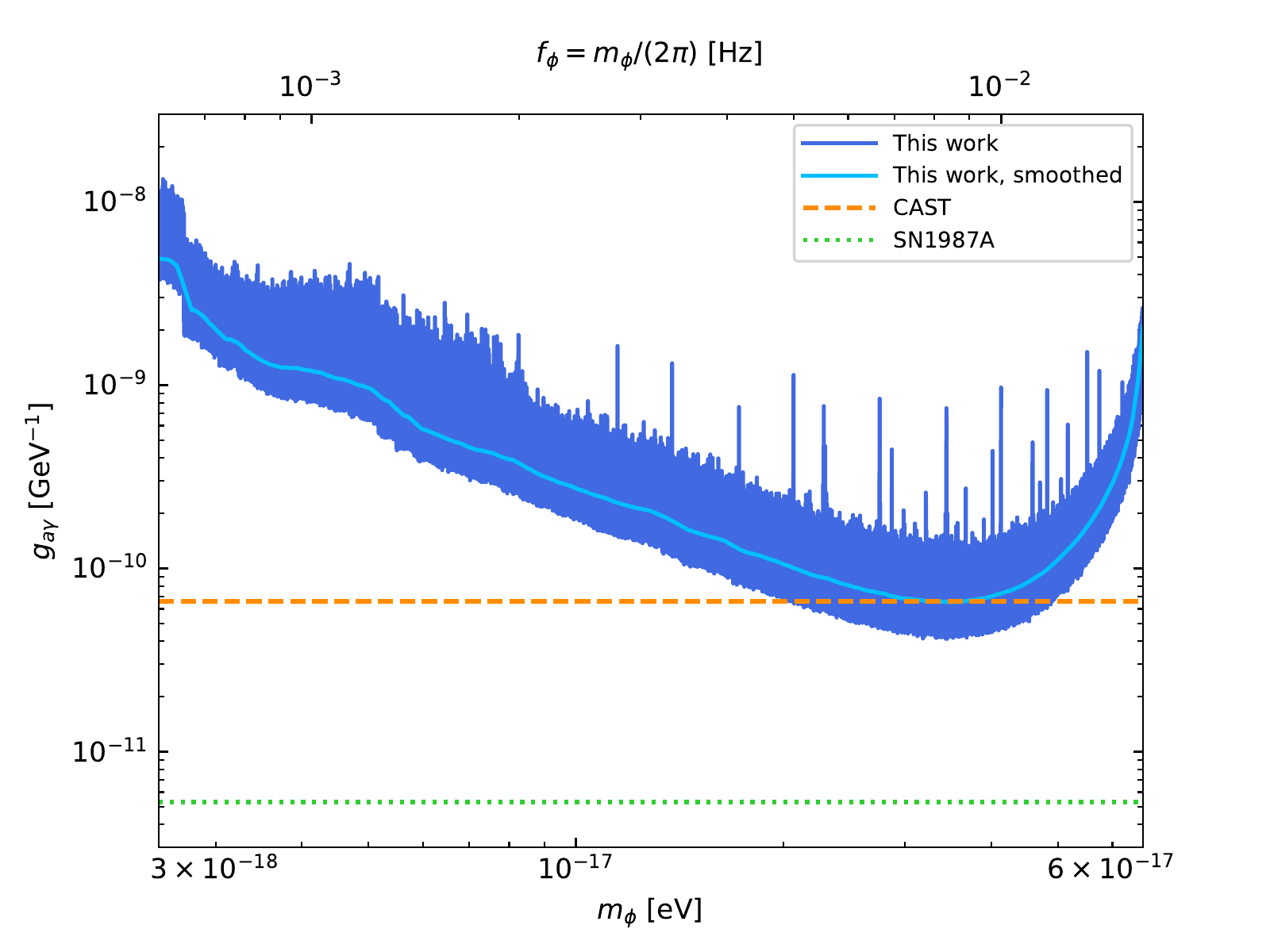}
\caption{\label{fig:limit}%
    The 95\%-credible exclusion limit on the axion-photon coupling $g_{a\gamma}$ based on a search of the SuperMAG dataset for the axion dark-matter magnetic-field signal \eqref{signal}, as summarized in \secref{analysis} and detailed in \appref{analysisdetails}.
    Our exclusion limit is shown as a function of the axion mass $m_a$ in solid dark blue and appears as a wide band due to the density of masses at which limits are plotted.
    The solid light blue line shows the sliding average of our exclusion limit over nearby frequencies.
    Our exclusion limit exhibits several narrow spikes (each at most a few frequency bins wide), which correspond to potential signal candidates.
    We investigate these candidates further in \appref{reevaluation} and show that none constitute robust evidence for dark matter.
    Also shown are existing limits from the CAST helioscope search for axions produced in the Sun~\cite{Anastassopoulos:2017ftl}~(dashed orange), and a constraint due to non-observation of a gamma-ray signal from axions in coincidence with SN1987A~\cite{Payez:2014xsa}~(dotted green).
    }
\end{figure*}

Based on this initial analysis, we identified 27 such candidates (some of which can be seen as narrow peaks above the dark blue exclusion band in \figref{limit}).
We then re-evaluated each na\"ive signal candidate for robustness, to test if it exhibits features of a physical axion dark-matter signal.
In particular, a physical axion dark-matter signal should be present for the entire duration of time over which SuperMAG has collected data, and should appear in the data from all stations across the globe.%
\footnote{\label{ftnt:projectionAndSpliting}%
    The manner in which our main analysis proceeds does attempt to project onto this global mode, but the exact fashion in which this is done leaves the main analysis vulnerable to a false-positive signal identification if a small subset of the stations exhibit a very large signal oriented in the appropriate direction.
    The point of the re-evaluation/robustness tests is to exclude this vulnerability.
    } %
Because of this, we partitioned the full SuperMAG dataset into four temporal subsets, consisting of data only from certain time periods, and four geographical subsets, consisting of data only from certain stations.
We re-performed our analysis on each of these subsets and searched for each na\"ive signal candidate to see if it re-appeared in the subset.
In particular, we checked if the analysis variables constructed from each data subset were consistent with the signal size implied by the original analysis of the full SuperMAG dataset, as characterized by the Bayesian posterior on $g_{a\gamma}$ constructed from the latter (see \appref{bayesian}).
We combined the results of all eight of these resampling checks and immediately rejected any candidates with a combined $p$-value of $p_\text{full}<0.01$ (see \appref{reevaluation}).

We note three candidates of remaining potential interest that we were unable to immediately reject based on this criterion: 
(1) one candidate at $f_a \approx 4.2\,\text{mHz}$ with a high ($6.7\sigma$) global significance that is in strong but not definitive tension ($0.01<p_\text{full}<0.05$) with the combined spatio-temporal robustness tests and also the geographical tests alone ($0.01<p_\text{geo}<0.05$); 
(2) one candidate at $f_a\approx5.5\,\text{mHz}$ with $3.6\sigma$ global significance that is however in strong tension with the temporal resampling checks ($0.01<p_\text{time}<0.05$); and 
(3) one candidate (actually a pair symmetrically arranged around the Nyquist frequency) at $f_a\approx8.3\,\text{mHz}$ with a low ($2.3\sigma$) global significance that is nevertheless consistent with all resampling checks.
While we do not consider these candidates to constitute strong and robust evidence for dark matter, they would require further work to definitively exclude.

With no robust axion dark-matter candidates identified, we set 95\%-credible exclusion limits (local significance) on the axion--photon coupling $g_{a\gamma}$ based on the Bayesian posterior on $g_{a\gamma}$ derived in our analysis.
\figref{limit} shows our exclusion limit on $g_{a\gamma}$ as a function of $m_a$, assuming that the axion is all of the dark matter and that $\rho_{\textsc{dm}} = 0.3\,\text{GeV/cm}^3$.
Also shown in \figref{limit} are limits on $g_{a\gamma}$ set by the CAST solar axion search~\cite{Anastassopoulos:2017ftl}, and limits based on non-observation of gamma-rays in coincidence with supernova SN1987A by the Gamma-Ray Spectrometer instrument on the Solar Maximum Mission satellite~\cite{Payez:2014xsa}.
The latter limit arises because axions produced via the Primakoff process in SN1987A would convert to gamma rays in the Milky Way's magnetic field.%
\footnote{\label{ftnt:magnetic_coherence}%
    It was recently pointed out~\cite{Carenza:2021alz} that accounting for the turbulent component of the Milky Way magnetic field complicates the computation of the conversion of astrophysical axions into gamma rays.
    Accounting for this turbulent component can change the axion--photon conversion probability by up to a factor of two.
} %
We do note that neither the CAST nor the SN1987A limits must assume that the axion is all of the dark matter.
Nevertheless, from \figref{limit}, it can be seen that our limits are competitive with CAST bounds in some mass ranges; they also rely on independent systematics.

A comment is in order on the mass dependence of our limits as shown in \figref{limit}: because our magnetic-field signal has $B\propto m_a a_0$ and because the root-mean-square (rms) value of $a_0 \propto \sqrt{\rho_{\textsc{dm}}} / m_a$, the rms amplitude of the magnetic field signal is independent of the axion mass once the rms axion field amplitude is normalized to the dark-matter density.
The $m_a$-dependence of our limits is thus driven by the underlying noise behavior of the SuperMAG dataset as a function of frequency [$f_a= m_a/(2\pi)$].
This contrasts with the dark-photon case~\cite{Fedderke:2021iqw,Fedderke:2021rys}, for which the signal amplitude itself also still depended linearly on the dark-photon mass, even after normalizing the dark-photon field amplitude to the dark-matter abundance (a well-known decoupling effect of the massless dark-photon limit).

In addition to the one-minute resolution dataset analyzed in this work, SuperMAG also maintains a one-second resolution dataset from a smaller number of stations.
A similar axion dark-matter search in this higher time-resolution dataset would allow for sensitivity to higher axion masses.
Given that our limit shown in \figref{limit} improves with increasing mass (owing to lower noise at higher frequencies in the one-minute SuperMAG dataset), we anticipate that the constraint from a search in the one-second resolution SuperMAG dataset could potentially outperform existing constraints from CAST in some mass range (and perhaps the SN1987A constraint, although this is less clear), assuming the noise in that dataset continues to behave similarly.
Additionally, a search in the higher time-resolution dataset would present an opportunity to re-evaluate the three na\"ive signal candidates of interest that we discussed above to see if they appear in that dataset.
We intend to undertake such a search in future work.

\section{Conclusion}
\label{sec:conclusion}

In this paper, we described a novel signature of ultralight axion dark matter with a coupling to photons $g_{a\gamma}$.
This signal is similar to that of dark-photon dark matter that was recently discussed in \citeR{Fedderke:2021rys} and searched for as described in \citeR{Fedderke:2021iqw}.
Namely, we pointed out that such an axion field converts off the quasi-static geomagnetic field of the Earth $\bm{B}_{0}$, to produce at ground-level all across the Earth's surface an observable magnetic-field signal $\bm{B}_a(t)$.
This signal oscillates at a frequency $f_a$ set by the axion mass $m_a \approx 2\pi f_a$, a fundamental physics parameter; moreover, it is narrowband in the sense that the bandwidth $\Delta f \sim v_{\textsc{dm}}^2 f_a \sim 10^{-6}f_a$, implying a long phase-coherence time for these oscillations.
The signal amplitude is $|\bm{B}_a| \sim g_{a\gamma} |\bm{B}_{0}| R \sqrt{\rho_{\textsc{dm}}}$ where $R$ is the radius of the Earth, implying that it is detectably large (note the appearance of $R$ here, and not some other length scale, such as the height of the atmosphere).
Finally, the signal has a global vectorial pattern that is set by the Earth's quasi-static geomagnetic field.

As such, this signal is an ideal candidate to be searched for using a network of globally distributed, terrestrial magnetic-field metrology stations.
There is an existing publicly available dataset of measurements of this type maintained by the SuperMAG Collaboration~\cite{Gjerloev:2009wsd,Gjerloev:2012sdg}, which consists of 50 years' worth of one-minute-resolution, three-axis magnetometer readings taken at 508 geographically dispersed stations in total (although not all stations report data at all times).
We made use of this dataset to search for our axion-induced magnetic-field signal in the axion mass range $2\times 10^{-18}\,\text{eV}\lesssim m_a \lesssim 7\times 10^{-17}\,\text{eV}$, constructing our analysis around projections of this large dataset onto a small number of vector spherical harmonic coefficients in which our signal is expected to appear.
Our search initially identified 27 na\"ive signal candidates in the data that at a global 95\% confidence level were inconsistent with a background-only hypothesis.
However, applying further robustness checks to test these candidates for spatial consistency and temporal uniformity, we definitively eliminated all but three of them.
The three candidates that were not definitively eliminated however still exhibited strong tension with (at least some subset of) our robustness tests, or had weak global significance.
As such, we do not consider any of them to be strong and robust signals of axion dark matter on the basis of this analysis.

Having dismissed all the anomalies in the data as either relatively weak and/or exhibiting of some defect and thus not robust, we turned to setting limits on the axion--photon coupling $g_{a \gamma}$. 
We made use of a Bayesian analysis procedure that folded in the effects of the stochastic fluctuations of the axion dark-matter field amplitude from one coherence time to the next.
Assuming that the axion is all of the DM, we set 95\%-credible upper limits (local significance) on the axion--photon coupling $g_{a \gamma}$ as a function of the axion mass $m_a$ in the same mass-range as our signal search; see \figref{limit}.
These limits are strongest for $3\times 10^{-18}\,\text{eV} \lesssim m_a \lesssim 4\times 10^{-18}\,\text{eV}$: smoothed over frequency-to-frequency fluctuations, the mean limit in this mass-range reaches $g_{a \gamma} \lesssim 6.5\times 10^{-11}\,\text{GeV}^{-1}$, which is comparable to limits on axions set by the CAST helioscope~\cite{Anastassopoulos:2017ftl}; they are however about an order of magnitude weaker than astrophysical limits set by observations of SN1987A~\cite{Payez:2014xsa} (but see also \citeR{Carenza:2021alz}).

The already impressive reach for this search, which has dramatically different systematics as compared to the other constraints in this axion mass range and is thus both competitive and complementary, could be further improved by future analysis of a higher resolution (one-second) dataset also maintained by the SuperMAG Collaboration.
Although this dataset has data from fewer stations and over a shorter total temporal duration as compared to the data analyzed in this work, if the decrease in the noise moving to higher frequencies that is evident in \figref{limit} persists also in that other dataset, it is possible that such an analysis could further probe for signals below existing CAST bounds at frequencies up to a factor of 60 higher than those searched in the present work. 
Moreover, this would afford the opportunity to revisit some of the weak or non-robust anomalies observed in this work for further analysis.
In future planned work that will be undertaken in collaboration with members of the SuperMAG Collaboration, we will apply the analysis techniques we have developed in this work and in \citeR[s]{Fedderke:2021rys,Fedderke:2021iqw} to this one-second resolution SuperMAG dataset.

\acknowledgments

This work was supported by the U.S.~Department of Energy (DOE), Office of Science, National Quantum Information Science Research Centers, Superconducting Quantum Materials and Systems Center (SQMS) under contract No.~DE-AC02-07CH11359. 
The work was also supported by the U.S.~National Science Foundation (NSF) Grant No.~PHY-2110388, Simons Investigator Grant No.~824870, NSF Grant No.~PHY-2014215, DOE HEP QuantISED Award No.~100495, and the Gordon and Betty Moore Foundation Grant No.~GBMF7946.

Some of the computing for this project was performed on the Sherlock cluster. 
We thank Stanford University and the Stanford Research Computing Center for providing computational resources and support that contributed to these research results.

The work of M.A.F.~was performed in part at the Aspen Center for Physics, which is supported by NSF Grant No.~PHY-1607611.

We gratefully acknowledge the SuperMAG Collaboration for maintaining and providing the database of ground magnetometer data that were analyzed in this work, and we thank Jesper W.~Gjerloev for helpful correspondence regarding technical aspects of the SuperMAG data.
SuperMAG receives funding from NSF Grant Nos.~ATM-0646323 and AGS-1003580, and NASA Grant No.~NNX08AM32G S03.

We acknowledge those who contributed data to the SuperMAG Collaboration: 
INTERMAGNET, Alan Thomson; 
CARISMA, PI Ian Mann; 
CANMOS, Geomagnetism Unit of the Geological Survey of Canada; 
The S-RAMP Database, PI K.~Yumoto and Dr.~K.~Shiokawa; 
The SPIDR database; AARI, PI Oleg Troshichev; 
The MACCS program, PI M.~Engebretson; 
GIMA; 
MEASURE, UCLA IGPP and Florida Institute of Technology; 
SAMBA, PI Eftyhia Zesta; 
210 Chain, PI K.~Yumoto; 
SAMNET, PI Farideh Honary; 
IMAGE, PI Liisa Juusola; 
Finnish Meteorological Institute, PI Liisa Juusola; 
Sodankylä Geophysical Observatory, PI Tero Raita; 
UiT the Arctic University of Norway, Troms\o\ Geophysical Observatory, PI Magnar G.~Johnsen; 
GFZ German Research Centre For Geosciences, PI J\"urgen Matzka; 
Institute of Geophysics, Polish Academy of Sciences, PI Anne Neska and Jan Reda; 
Polar Geophysical Institute, PI Alexander Yahnin and Yarolav Sakharov; 
Geological Survey of Sweden, PI Gerhard Schwarz; 
Swedish Institute of Space Physics, PI Masatoshi Yamauchi; 
AUTUMN, PI Martin Connors; 
DTU Space, Thom Edwards and PI Anna Willer; 
South Pole and McMurdo Magnetometer, PIs Louis J.~Lanzarotti and Alan T.~Weatherwax; 
ICESTAR; 
RAPIDMAG; 
British Antarctic Survey; 
McMac, PI Dr.~Peter Chi; 
BGS, PI Dr.~Susan Macmillan; 
Pushkov Institute of Terrestrial Magnetism, Ionosphere and Radio Wave Propagation (IZMIRAN); 
MFGI, PI B.~Heilig; 
Institute of Geophysics, Polish Academy of Sciences, PI Anne Neska and Jan Reda; 
University of L’Aquila, PI M.~Vellante; 
BCMT, V.~Lesur and A.~Chambodut; 
Data obtained in cooperation with Geoscience Australia, PI Marina Costelloe; 
AALPIP, co-PIs Bob Clauer and Michael Hartinger; 
SuperMAG, PI Jesper W.~Gjerloev; 
data obtained in cooperation with the Australian Bureau of Meteorology, PI Richard Marshall.

We thank INTERMAGNET for promoting high standards of magnetic observatory practice~\cite{INTERMAGNETwebsite}.

\appendix

\section{Vector spherical harmonics}
\label{app:vectorSphericalHarmonics}

This appendix, which defines the VSH conventions used in this work, is reproduced from \citeR{Fedderke:2021rys} with minor modifications for the convenience of the reader.

The VSH are defined in terms of the scalar spherical harmonics $Y_{\ell m}$ by the relations,
\begin{align}\label{eq:VSHdef}
    \bm{Y}_{\ell m} &= Y_{\ell m}\rhat, &
    \bm{\Psi}_{\ell m} &= r\bm{\nabla} Y_{\ell m}, &
    \bm{\Phi}_{\ell m} &= \bm{r}\times\bm{\nabla} Y_{\ell m},
\end{align}
where $\rhat$ is the radial unit vector.
Thus $\bm{Y}_{\ell m}$ points radially, while $\bm{\Psi}_{\ell m}$ and $\bm{\Phi}_{\ell m}$ point tangentially.
Some of their relevant properties (and our phase conventions) are
\begin{align}
    \bm{Y}_{\ell,-m}&=(-1)^m\bm{Y}_{\ell m}^*,\label{eq:Yminus}\\
    \bm{\Psi}_{\ell,-m}&=(-1)^m\bm{\Psi}_{\ell m}^*,\label{eq:Psiminus}\\
    \bm{\Phi}_{\ell,-m}&=(-1)^m\bm{\Phi}_{\ell m}^*, \label{eq:Phiminus}
\end{align}
\begin{align}
    \bm{Y}_{\ell m}\cdot\bm{\Psi}_{\ell m}&=\bm{Y}_{\ell m}\cdot\bm{\Phi}_{\ell m}=\bm{\Psi}_{\ell m}\cdot\bm{\Phi}_{\ell m}=0,
\end{align}
\begin{align}
    \int d\Omega\,\bm{Y}_{\ell m}\cdot \bm{Y}_{\ell'm'}^*&=\delta_{\ell\ell'}\delta_{mm'},\label{eq:Y_orthog}\\
    \int d\Omega\,\bm{\Psi}_{\ell m}\cdot\bm{\Psi}_{\ell'm'}^*&=\int d\Omega~\Phi_{\ell m}\cdot\Phi_{\ell'm'}^*\nonumber \\
	&=\ell(\ell+1)\delta_{\ell\ell'}\delta_{mm'},\label{eq:PhiPsi_orthog}\\
    \int d\Omega\,\bm{Y}_{\ell m}\cdot\bm{\Psi}_{\ell'm'}^*&=\int d\Omega\,\bm{Y}_{\ell m}\cdot\bm{\Phi}_{\ell'm'}^*\nonumber \\
	&=\int d\Omega\,\bm{\Psi}_{\ell m}\cdot\bm{\Phi}_{\ell'm'}^*=0.\label{eq:cross_orthog}
\end{align}

For any radially dependent function $f(r)$, the gradient of the scalar spherical harmonics can be related to the VSH by
\begin{align}
    \nabla\lb(fY_{\ell m}\rb)=\frac{df}{dr}\bm{Y}_{\ell m}+\frac{f}{r}\bm{\Psi}_{\ell m}.\label{eq:gradient}
\end{align}
Additionally, the divergences and curls of the VSH are given by
\begin{align}
    \nabla \cdot \lb(f\bm{Y}_{\ell m}\rb)&=\lb( \frac{df}{dr} + \frac{2f}{r} \rb) Y_{\ell m},\label{eq:Ydiv}\\
    \nabla \cdot \lb(f\bm{\Psi}_{\ell m}\rb)&= -\ell(\ell+1) \frac{f}{r} Y_{\ell m},\label{eq:Psidiv}\\
    \nabla \cdot \lb(f\bm{\Phi}_{\ell m}\rb)&=0,\label{eq:Phidiv}\\[3ex]
    \nabla\times\lb(f\bm{Y}_{\ell m}\rb)&=-\frac fr\bm{\Phi}_{\ell m},\label{eq:Ycurl}\\
    \nabla\times\lb(f\bm{\Psi}_{\ell m}\rb)&=\lb(\frac{df}{dr}+\frac fr\rb)\bm{\Phi}_{\ell m},\label{eq:Psicurl}\\
    \nabla\times\lb(f\bm{\Phi}_{\ell m}\rb)&=-\frac{\ell(\ell+1)f}r\bm{Y}_{\ell m}-\lb(\frac{df}{dr}+\frac fr\rb)\bm{\Psi}_{\ell m},\label{eq:Phicurl}
\end{align}
with the Laplacians then being
\begin{align}
    \nabla^2\lb(f\bm{Y}_{\ell m}\rb)
        &=\lb(\frac1{r^2}\frac d{dr}\lb(r^2\frac{df}{dr}\rb)-\frac{(\ell(\ell+1)+2)f}{r^2}\rb)\bm{Y}_{\ell m}
        \nl\quad +\frac{2f}{r^2}\bm{\Psi}_{\ell m},\label{eq:Ylaplace}\\
    \nabla^2\lb(f\bm{\Psi}_{\ell m}\rb)
        &=\lb(\frac1{r^2}\frac d{dr}\lb(r^2\frac{df}{dr}\rb)-\frac{\ell(\ell+1)f}{r^2}\rb)\bm{\Psi}_{\ell m}
        \nl\quad +\frac{2\ell(\ell+1)f}{r^2}\bm{Y}_{\ell m},\label{eq:Psilaplace}\\
    \nabla^2\lb(f\bm{\Phi}_{\ell m}\rb)
        &=\lb(\frac1{r^2}\frac d{dr}\lb(r^2\frac{df}{dr}\rb)-\frac{\ell(\ell+1)f}{r^2}\rb)\bm{\Phi}_{\ell m}.\label{eq:Philaplace}
\end{align}

\section{Detailed signal calculation}
\label{app:detailedcalculation}

In this appendix, we derive in more detail the axion dark-matter induced magnetic-field signal, \eqref{signal}, that would be measured in the lower-atmospheric air gap just above the surface of the Earth.
The calculation in this appendix closely follows the calculations in Secs.~III\,B and~III\,C of \citeR{Fedderke:2021rys}, but using the effective current given by \eqref{Eartheff_current}.
As in \citeR{Fedderke:2021rys}, we will derive the axion dark-matter signal using two different models of the near-Earth conductivity environment.
In the first, we will take the inner and outer boundaries of the lower-atmospheric air gap to be spherical perfectly conducting shells at $r=R$ (corresponding to the Earth) and $r=R+L$ (corresponding to the ionosphere), respectively, with the lower-atmospheric region separating them assumed to be vacuum.
In this case, the magnetic-field signal at $r=R$, to leading order in $m_a R$, will be exactly \eqref{signal}.
In the second model, we will still take the inner boundary of the lower-atmospheric air-gap to be a spherical Earth (this is accurate to 0.3\%~\cite{WGS84}), but we will allow the outer boundary to have an arbitrary shape; this corresponds to the scenario where the outer boundary is the interplanetary medium.
In this case, \eqref{signal} will give only the leading $\bm\Phi_{\ell m}$ contribution to the magnetic-field signal at $r=R$; in general, other $\bm Y_{\ell m}$ and $\bm\Psi_{\ell m}$ contributions may also be present, but these contributions may depend on the details of the outer boundary.

In either model, we can decompose the electric field in the vacuum region into two contributions (cf.~Eqs.~(14)--(16) of \citeR{Fedderke:2021rys}),
\begin{align}
    \bm{E}=\bm{E}_\text{inh}+\bm{E}_\text{hom},
\end{align}
where $\bm{E}_\text{inh}$ is chosen to satisfy
\begin{align}
    (\nabla^2-\partial_t^2)\bm{E}_\text{inh}=\partial_t\bm{J}_\text{eff},
    \label{eq:Einh}
\end{align}
and $\bm{E}_\text{hom}$ is chosen so that $\bm{E}$ fulfills the boundary conditions on the full solution, while satisfying
\begin{align}
    (\nabla^2-\partial_t^2)\bm{E}_\text{hom}=0.
    \label{eq:Ehom}
\end{align}
Both contributions must also satisfy
\begin{align}
    \nabla\cdot\bm{E}_\text{inh/hom}=0.
\end{align}

Because of this last criterion, both contributions must be composed of `transverse electric' (TE) and `transverse magnetic' (TM) modes, whose electric fields are of the form
\begin{align}
    \label{eq:ETE}
    \bm E_\text{TE}&\equiv\sum_{\ell,m}f_{\ell m}(m_a r)\bm\Phi_{\ell m}e^{-im_a t},\\
    \bm E_\text{TM}&\equiv\sum_{\ell,m}\frac1{m_a}\nabla\times\left(g_{\ell m}(m_a r)\bm\Phi_{\ell m}\right)e^{-im_a t}\nonumber\\
    &=\sum_{\ell,m}\left(\begin{array}{l}-\dfrac{\ell(\ell+1)g_{\ell m}(m_a r)}{m_a r}\bm Y_{\ell m}\\[2ex]-\left(g'_{\ell m}(m_a r)+\dfrac{g_{\ell m}(m_a r)}{m_a r}\right)\bm\Psi_{\ell m}\end{array}\right)\nl \qquad\times e^{-im_a t},
    \label{eq:ETM}
\end{align}
for some scalar functions $f_{\ell m}$ and $g_{\ell m}$ to be determined below.
We denote the TE and TM contributions to $\bm E_\text{inh}$ by $\bm E_\text{inh,TE}$ and $\bm E_\text{inh,TM}$, with relevant scalar functions $f_{\text{inh},\ell m}$ and $g_{\text{inh},\ell m}$ (and likewise for $\bm E_\text{hom}$).

Comparing \eqref{Eartheff_current} with the forms of the RHS of \eqref[s]{ETE} and (\ref{eq:ETM}), it can be shown straightforwardly that $\bm J_\text{eff}$ has the same form as the RHS of \eqref{ETM}, with
\begin{align}
    g_{\text{eff},\ell m}(x)=-ig_{a\gamma}m_aa_0C_{\ell m}\cdot\frac{x_0^{\ell+2}}{\ell x^{\ell+1}},
\end{align}
where $x\equiv m_a r$ and $x_0\equiv m_a R$.
Substituting \eqref{ETM} into \eqref{Einh} and making use of the VSH Laplacian properties \eqrefRange{Ylaplace}{Philaplace}, it follows that
\begin{widetext}
\begin{align}
    x^2g''_{\text{inh},\ell m}(x)+2xg'_{\text{inh},\ell m}(x)+(x^2-\ell(\ell+1))g_{\text{inh},\ell m}(x)=-\frac{ix^2g_{\text{eff},\ell m}(x)}{m_a}=-g_{a\gamma}a_0C_{\ell m}\cdot\frac{x_0^{\ell+2}}{\ell x^{\ell-1}}.
\end{align}
These equations, one for each $(\ell,m)$ pair, are solved by%
\interfootnotelinepenalty=10000
\footnote{\label{ftnt:inspection}%
    This solution can be read off by inspection after noting that $x^2(x^{-(l+1)})'' + 2x (x^{-(l+1)})' - \ell(\ell+1)(x^{-(l+1)}) = 0$.
    } %
\interfootnotelinepenalty=100
\begin{align}
    g_{\text{inh},\ell m}=-g_{a\gamma}a_0C_{\ell m}\cdot\frac{x_0^{\ell+2}}{\ell x^{\ell+1}}.
\end{align}
Likewise it can be readily seen that \eqref{Ehom} implies
\begin{align}
    x^2g''_{\text{hom},\ell m}(x)+2xg'_{\text{hom},\ell m}(x)+(x^2-\ell(\ell+1))g_{\text{hom},\ell m}(x)=0,
\end{align}
so that $g_{\text{hom},\ell m}$ are linear combinations of spherical Bessel functions
\begin{align}
    g_{\text{hom},\ell m}(x)=A_{\ell m}j_\ell(x)+B_{\ell m}x_0^{2\ell+1}y_\ell(x),
\end{align}
where we have extracted a factor of $x_0^{2\ell+1}$ from $B_{\ell m}$ so that $A_{\ell m}$ and $B_{\ell m}$ have the same power-counting in an expansion in $x_0$.

\subsection{Spherical boundary conditions}
\label{app:sphericalboundary}

Now to solve for $A_{\ell m}$ and $B_{\ell m}$ and obtain the full solution, we must fix boundary conditions.
Let us first consider the near-Earth conductivity model with spherical boundaries, where the inner boundary lies at $r=R$ and the outer boundary lies at $r=R+L$.
This means that the $\bm\Psi_{\ell m}$ components of the full electric field must vanish at these radii (as they are parallel to the boundaries).
This implies the boundary conditions,
\begin{align}
    \label{eq:boundary1}
    g'_{\text{hom},\ell m}(m_a R)+\frac{g_{\text{hom},\ell m}(m_a R)}{m_a R}&=-g_{a\gamma}a_0C_{\ell m},\\
    g'_{\text{hom},\ell m}(m_a(R+L))+\frac{g_{\text{hom},\ell m}(m_a(R+L))}{m_a(R+L)}&=-g_{a\gamma}a_0C_{\ell m}\left(\frac R{R+L}\right)^{\ell+2}.
    \label{eq:boundary2}
\end{align}
Let us assume that the axion Compton wavelength is much larger than the largest physical length-scales in the problem: $m_a R,m_a(R+L)\ll1$.
We can then employ the small-$x$ expansion of the spherical Bessel functions
\begin{align}
    j_\ell(x)&\sim\frac{x^\ell}{(2\ell+1)!!}+\mathcal O(x^{\ell+2}),\\
    y_\ell(x)&\sim-(2\ell-1)!!x^{-\ell-1}+\mathcal O(x^{-\ell+1}),
    \label{eq:BesselY}
\end{align}
where $n!!=n\cdot(n-2)\cdot(n-4)\cdots k_n$, where $k_n$ is the smallest positive integer with the same parity as $n$.
The leading terms in \eqrefRange{boundary1}{boundary2} in this expansion are
\begin{align}
    \frac{\ell+1}{(2\ell+1)!!}A_{\ell m}(m_a R)^{\ell-1}+\ell(2\ell-1)!!B_{\ell m}(m_a R)^{\ell-1}&=-g_{a\gamma}a_0C_{\ell m},\\
    \frac{\ell+1}{(2\ell+1)!!}A_{\ell m}(m_a(R+L))^{\ell-1}+\ell(2\ell-1)!!B_{\ell m}(m_a R)^{\ell-1}\left(\frac R{R+L}\right)^{\ell+2}&=-g_{a\gamma}a_0C_{\ell m}\left(\frac R{R+L}\right)^{\ell+2}.
\end{align}
This is solved by $A_{\ell m}=0$ and
\begin{align}
    B_{\ell m}=-\frac{g_{a\gamma}a_0C_{\ell m}}{\ell(2\ell-1)!!(m_a R)^{\ell-1}},
\end{align}
with corrections appearing at higher orders in $m_a R$ and $m_a(R+L)$; note that this expansion has \emph{not} assumed anything as to the relative sizes of $R$ and $L$. 
Interestingly, this is in a sense the opposite outcome to that of the dark-photon calculation in \citeR{Fedderke:2021rys}, where the $B_{\ell m}$ were the coefficients that vanished to leading order.

As in the dark-photon case, all contributions to the full electric field vanish everywhere in the cavity to order $\mathcal O(x_0^2)$,
\begin{align}
    \bm E_{\ell m}&=\bm E_{\text{inh},\ell m}+\bm E_{\text{hom},\ell m}\\
    &=\left(\begin{array}{l}-\left(\dfrac{\ell(\ell+1)g_{\text{inh},\ell m}(x)}x+\dfrac{\ell(\ell+1)g_{\text{hom},\ell m}(x)}x\right)\bm Y_{\ell m}\\[2ex]-\left(g'_{\text{inh},\ell m}(x)+\dfrac{g_{\text{inh},\ell m}(x)}x+g'_{\text{hom},\ell m}(x)+\dfrac{g_{\text{hom},\ell m}(x)}x\right)\bm\Psi_{\ell m}\end{array}\right)\times e^{-im_a t}\\
    &=g_{a\gamma}a_0C_{\ell m}\left(\begin{array}{l}(\ell+1)\left(\left(\dfrac{x_0}x\right)^{\ell+2}+\dfrac{x_0^{\ell+2}y_\ell(x)}{(2\ell-1)!!x}\right)\bm Y_{\ell m}\\[2ex]-\left(\left(\dfrac{x_0}x\right)^{\ell+2}-\dfrac{x_0^{\ell+2}}{\ell(2\ell-1)!!}\left(y'_\ell(x)+\dfrac{y_\ell(x)}x\right)\right)\bm\Psi_{\ell m}\end{array}\right)\times e^{-im_a t}+\mathcal O(x_0^2)=\mathcal O(x_0^2),
    \label{eq:Evanish}
\end{align}
where at the last equality, we used the expansion \eqref{BesselY} in the $(\,\cdots)$ bracket, which causes the leading $(x_0/x)^{l+2}$ term to vanish leaving a leading correction at $\mathcal{O}(x_0^2(x_0/x)^{l})$; since $x\geq x_0$, the $(\,\cdots)$ bracket is parametrically $\mathcal{O}(x_0^2)$ too.
The magnetic-field contributions, on the other hand, can be derived by Faraday's law, \eqref{Faradays_Law}.
Because the inhomogeneous contributions have no associated magnetic field
\begin{align}
    \bm B_{\text{inh},\ell m}&=-\frac i{m_a}\nabla\times\bm E_{\text{inh},\ell m}=i\left(g''_{\text{inh},\ell m}(x)+\frac{2g'_{\text{inh},\ell m}(x)}{x}-\frac{\ell(\ell+1)g_{\text{inh},\ell m}(x)}{x^2}\right)\bm\Phi_{\ell m}e^{-im_a t}\\
    &=\left(-ig_{\text{inh},\ell m}(x)+\frac{g_{\text{eff},\ell m}(x)}{m_a}\right)\bm\Phi_{\ell m}e^{-im_a t}=0,
\end{align}
the full magnetic field comes solely from the homogeneous contributions:
\begin{align}
    \bm B=\bm B_\text{hom}=-\frac i{m_a}\nabla\times\bm E_\text{hom}=-i\sum_{\ell,m}g_{\text{hom},\ell m}\bm\Phi_{\ell m}e^{-im_a t}=-ig_{a\gamma}a_0m_a R\sum_{\ell,m}\frac{C_{\ell m}}\ell\left(\frac Rr\right)^{\ell+1}\bm\Phi_{\ell m}e^{-im_a t},
    \label{eq:appendix_signal}
\end{align}
which agrees with \eqref{signal}.
\end{widetext}

\subsection{Aspherical boundary conditions}
\label{app:asphericalboundary}

Now consider the case where the boundaries have arbitrary shape (but we are still interested in the magnetic field at $r=R$).
As shown in \eqref{Evanish}, the electric field in the above solution vanishes \emph{everywhere} in the cavity in \emph{all directions} up to corrections at $\mathcal O[(m_a R)^2]$.
This means that one could perturb the boundary shape in any arbitrary fashion, so long as the largest length-scale associated with boundary remains smaller than the Compton wavelength of the axion, and the correct electric-field boundary condition would still be satisfied on that perturbed boundary up to corrections at $\mathcal O[(m_a R)^2]$. 
Therefore, regardless of the boundary conditions, the above solution will remain the correct solution for the homogeneous electric field up to corrections at order $\mathcal O[(m_a R)^2]$. 
It then only remains to determine what this implies for the magnetic-field solution.

There is a power-counting argument for the magnetic field solution that can be followed and which proceeds similarly to the argument in Sec.~III\,C of \citeR{Fedderke:2021rys}.
Namely, if we Taylor expand $\bm E_\text{hom,TE}$ in $x_0=m_a R$ as
\begin{align}
    \bm E_\text{hom,TE}=\sum_{n=0}^\infty x_0^n\bm E^{(n)}_\text{hom,TE},
\end{align}
with
\begin{align}
    \bm E^{(n)}_\text{hom,TE}\equiv\sum_{\ell,m}f^{(n)}_{\ell m}(\xi)\bm\Phi_{\ell m}e^{-im_a t},
    \label{eq:EnTE}
\end{align}
for some scalar functions $f^{(n)}_{\ell m}(\xi)$, where $\xi=r/R$, then it can be shown using Faraday's law \eqref{Faradays_Law} and the VSH curl identities \eqrefRange{Ycurl}{Phicurl} that the associated magnetic field has the expansion
\begin{align}
    \bm B_\text{hom,TE}=\sum_{n=0}^\infty x_0^n\bm B_\text{hom,TE}^{(n+1)},
\end{align}
with
\begin{align}
    \bm B^{(n)}_\text{hom,TE}\equiv-i\sum_{\ell,m}\nabla_\xi\times\left(f^{(n)}_{\ell m}(\xi)\bm\Phi_{\ell m}\right)e^{-im_a t},
    \label{eq:BnTE}
\end{align}
for some scalar functions $g^{(n)}_{\ell m}$, where $\nabla_\xi$ denotes that the radial co-ordinate in the derivative is taken to be $\xi$.
Importantly, $\bm B^{(n)}_\text{hom,TE}$ is determined by $\bm E^{(n)}_\text{hom,TE}$, so that the former vanishes everywhere within the bounded volume if the latter does. 
Likewise the expansions for the TM contributions are related by
\begin{align}
    \bm E_\text{hom,TM}&=\sum_{n=0}^\infty x_0^n\bm E_\text{hom,TM}^{(n+1)},\\
    \bm B_\text{hom,TM}&=\sum_{n=0}^\infty x_0^n\bm B_\text{hom,TM}^{(n)},
\end{align}
where
\begin{align}
    \label{eq:EnTM}
    \bm E^{(n)}_\text{hom,TM}&\equiv\sum_{\ell,m}\nabla_\xi\times\left(g^{(n)}_{\ell m}(\xi)\bm\Phi_{\ell m}\right)e^{-im_a t},\\
    \bm B^{(n)}_\text{hom,TM}&\equiv-i\sum_{\ell,m}g^{(n)}_{\ell m}(\xi)\bm\Phi_{\ell m}e^{-im_a t}.
    \label{eq:BnTM}
\end{align}
Thus the full homogeneous electric and magnetic fields can be expanded as
\begin{align}
    \bm E_\text{hom}&=\sum_{n=0}^\infty x_0^n\bm E_\text{hom}^{(n)},\\
    \bm B_\text{hom}&=\sum_{n=0}^\infty x_0^n\bm B_\text{hom}^{(n)},
\end{align}
where
\begin{align}
    \bm E_\text{hom}^{(n)}&=\bm E_\text{hom,TE}^{(n)}+\bm E_\text{hom,TM}^{(n+1)},\\
    \bm B_\text{hom}^{(n)}&=\bm B_\text{hom,TE}^{(n+1)}+\bm B_\text{hom,TM}^{(n)}.
\end{align}

As noted above, the homogeneous electric-field solution in the non-spherical case will be precisely the same as in the spherical case up to additional corrections at $\mathcal O(x_0^2)$.
Therefore, $\bm E_\text{hom}^{(0)}$ is as in the spherical case, and $\bm E_\text{hom}^{(1)}$ vanishes.
Because the spherical solution is entirely TM, then $\bm E_\text{hom}^{(0)}=\bm E_\text{hom,TM}^{(1)}$ and $\bm E_\text{hom,TE}^{(0)}=0$.
Moreover because $\bm E_\text{hom}^{(1)}$ vanishes, then $\bm E_\text{hom,TE}^{(1)}=0$ as well.
Because of the relations between $\bm E_\text{hom,TE/TM}^{(n)}$ and $\bm B_\text{hom,TE/TM}^{(n)}$ outlined in \eqref[s]{EnTE} and (\ref{eq:BnTE}) and \eqref[s]{EnTM} and (\ref{eq:BnTM}), this also implies that $\bm B_\text{hom,TE}^{(0)}=\bm B_\text{hom,TE}^{(1)}=0$ and $\bm B_\text{hom,TM}^{(1)}$ is given by the solution in the spherical case \eqref{appendix_signal} [or \eqref{signal}].
Therefore,
\begin{align}
    \bm B_\text{hom}^{(0)}=\bm B_\text{hom,TE}^{(1)}+\bm B_\text{hom,TM}^{(0)}=0,
\end{align}
and the leading order magnetic field will be
\begin{align}
    \bm B^{(1)}=\bm B_\text{hom}^{(1)}=\bm B_\text{hom,TE}^{(2)}+\bm B_\text{hom,TM}^{(1)},
    \label{eq:B1}
\end{align}
While it is clear from \eqref{B1} that $\bm B_\text{hom,TE}^{(2)}$ can give a leading-order contribution to the magnetic field, it will contribute to different VSH than $\bm B_\text{hom,TM}^{(1)}$.
Namely TE contributions to the magnetic field are comprised of $\bm Y_{\ell m}$ and $\bm\Psi_{\ell m}$ modes, while TM contributions are comprised of $\bm\Phi_{\ell m}$ modes.
Therefore, \eqref{signal} [or \eqref{appendix_signal}] indeed gives the correct leading order $\bm\Phi_{\ell m}$ contributions to the axion dark-matter magnetic-field signal, regardless of the boundary shape (so long as the largest physical scale in the problem is shorter than the axion Compton wavelength).

\section{Analysis details}
\label{app:analysisdetails}

In this appendix, we explain the details of the signal search whose results are summarized in \secref{analysis}.
The analysis in this work proceeds similarly to the dark-photon analysis described in \citeR{Fedderke:2021iqw}.
Here we therefore focus on the aspects of the axion analysis described in this work that differ from the dark-photon analysis, and we refer the reader to \citeR{Fedderke:2021iqw} for the parts of the analysis that are identical.

The analysis described in this appendix searches global magnetometer data maintained by the SuperMAG Collaboration~\cite{Gjerloev:2009wsd,Gjerloev:2012sdg,SuperMAGwebsite} for the signal \eqref{signal}.
The SuperMAG dataset that we analyze consists of time series of three-axis magnetic-field measurements from each of 508 stations which together cover a 50-year-long time period from the beginning of 1970 to the end of 2019.
Since many of the stations began reporting data later than the beginning of 1970, shut down prior to the end of 2019, or underwent periods of inactivity, the time series that each station reports is not continuous over the entire 50-year duration%
\footnote{\label{ftnt:duration}%
    For technical reasons related to the number of independent stations' measurements required to perform the analysis for a vectorial dark-photon dark-matter signal, the analysis in \citeR{Fedderke:2021iqw} restricted its attention to the data taken from the beginning of 1972 to the end of 2019; see footnotes 7 and 39 of \citeR{Fedderke:2021iqw}.
    These reasons do not apply to the analysis in this work, as only one active station is required to produce the two linearly independent time series $X^\theta$ and $X^\phi$ required to perform the analysis for the scalar axion dark-matter search.
    Therefore, the present analysis also utilizes the data from the years 1970 and 1971.
} %
of the SuperMAG dataset.
We denote the geographic coordinates of station $i$ by $\Omega_i=(\theta_i,\phi_i)$ and the three-axis magnetic-field measurement it reports at time $t_j$ by $\bm B_i(t_j) \equiv B_{i}^{\theta}(t_j) \bm{\hat{\theta}}_i +  B_{i}^{\phi}(t_j) \bm{\hat{\phi}}_i$.%
\footnote{\label{ftnt:SuperMAGaxes}%
    SuperMAG reports the magnetic-field measurements from each station in locally defined coordinates, oriented along Local Magnetic North and Local Magnetic East.
    Using the IGRF model, these measurements can be rotated on a station-by-station basis to globally defined coordinates, oriented along True Geographic North and True Geographic East.
    In what follows, we will work solely with the measurements in these geographic coordinates.
    In particular, we denote the components of $\bm B_i(t_j)$ by $B_i^\theta(t_j)$ (oriented towards geographic South) and $B_i^\phi(t_j)$ (oriented towards geographic East).
    The vertical component of $\bm B_i(t_j)$ will not be relevant for our analysis.
    See Sec.~III\,B of \citeR{Fedderke:2021iqw} for more details on the SuperMAG coordinate systems and the rotations required to achieve field measurements aligned to geographic co-ordinates.
} %
We also denote the set of sampling times at which station $i$ reports valid measurements by $\mathcal T_i$.
Importantly, $\mathcal T_i$ differs between stations, making a straightforward station-by-station analysis difficult.

Our search for an axion signal at frequency $f_a=m_a/2\pi$ (assuming $v_{\textsc{dm}}=0$; see footnote 34 in \citeR{Fedderke:2021iqw}) proceeds roughly as follows.
We combine the $B_i^\theta(t_j)$ measurements from all 508 stations into one time series $X^\theta(t_j)$, based on the $\thetahat$-component of the signal in \eqref{signal}, and we likewise combine the $B_i^\phi(t_j)$ measurements into another time series $X^\phi(t_j)$ based on the $\bm{\hat{\phi}}$-component of the signal.
We then partition each time series into segments $X^\theta_k(t_j)$ and $X^\phi_k(t_j)$ of duration roughly equal to the coherence time $T_{\text{coh}}$ of the axion dark-matter field (which depends on $f_a$).
We Fourier transform each segment to find $\tilde X^\theta_k(f_a)$ and $\tilde X^\phi_k(f_a)$, and combine them into a two-dimensional vector $\vec X_k$.

These $\vec X_k$ are the primary variables that we use below to construct the likelihood function for our analysis.
As such, we must compute both their expectation values under the signal hypothesis and variances under the background-only hypothesis.
Their expectation values under the signal hypothesis are computed by performing the same time-series combination on the signal \eqref{signal} as we performed on the SuperMAG data.
Their variances under the background-only hypothesis are computed by a data-driven noise estimation procedure, identical to the one outlined in \citeR{Fedderke:2021iqw}.

With the statistics of $\vec X_k$ computed, we construct a likelihood function under the assumption of a signal with coupling $g_{a\gamma}$. Assuming an objective Jeffreys prior on $g_{a\gamma}$, we use this likelihood in a Bayesian analysis framework to compute the posterior for $g_{a\gamma}$. 
Finally, this posterior is used to set 95\%-credible upper limits on $g_{a\gamma}$.

Of course, setting exclusion limits on a parameter should not be the main goal of a signal search; logically, the prior and more interesting question is whether the data support the inference of a non-zero signal above background.
Therefore, in addition to setting exclusion limits, we search for na\"ive signal candidates.
We perform this part of the analysis in a frequentist fashion, searching for isolated frequencies at which the data variables $\vec X_k$ are inconsistent with the absence of a signal as determined using the likelihood function under the assumption of $g_{a\gamma}=0$.
We identify 27 such candidate signals in our analysis that were inconsistent with the null hypothesis at 95\% global significance. 
We re-evaluate each such identified na\"ive signal candidate to test it for robustness: i.e., temporal consistency and spatial uniformity, which are required properties of a physical axion dark-matter signal.
These tests consist of splitting the full SuperMAG dataset into a number of smaller data subsets, either by restricting the temporal duration of the data to create temporally disjoint subsets, or by restricting the stations whose data are utilized to create sets of data recorded by disjoint sets of stations.
We check if the $X_k$ computed from the data subsets are consistent with the posterior on $g_{a\gamma}$ derived from the Bayesian analysis of the full dataset.

The subsections of this appendix will follow the above structure.
Namely, \appref{timeseries} will discuss the construction of the $X^\theta$ and $X^\phi$ time series and the $\vec X_k$ variables; \appref{Xkstats} will compute the expectation and variances of the $\vec X_k$ variables under the appropriate hypotheses; \appref{bayesian} will derive the likelihood function for the $\vec X_k$ variables and use it to set exclusion limits on $g_{a\gamma}$; and finally, \appref{reevaluation} will identify the na\"ive signal candidates in our analysis and test them for robustness.

\subsection{Time series construction}
\label{app:timeseries}

In \citeR{Fedderke:2021iqw}, five time series $X^{(n)}$ were constructed based on the five distinct $\thetahat$- and $\phihat$- components of the $\bm\Phi_{1m}$ modes.
Here we construct our time series in a similar manner, but instead only require two time series $X^\theta$ and $X^\phi$ based on the components of the signal \eqref{signal}.
In particular, we define%
\footnote{\label{ftnt:Xunits}%
    We note that since $C_{\ell m}$ has units of nT, then $X^\theta$ and $X^\phi$ in this work have units of $(\text{nT})^2$.
    This is in contrast to \citeR{Fedderke:2021iqw}, where the $X^{(n)}$ time series have units of nT.
} %
\begin{align}
    \label{eq:Xtheta}
    X^\theta(t_j) &= \frac1{W^\theta(t_j)}\sum_{\{i|t_j\in\mathcal T_i\}}w_i^\theta(t_j)B_i^\theta(t_j)\nl \qquad\qquad \qquad\quad\times \sum_{\ell,m}\frac{C_{\ell m}(t_j)}\ell\Phi_{\ell m}^\theta(\Omega_i),\\
    X^\phi(t_j) &= \frac1{W^\phi(t_j)}\sum_{\{i|t_j\in\mathcal T_i\}}w_i^\phi(t_j)B_i^\phi(t_j)\nl \qquad\qquad \qquad\quad\times \sum_{\ell,m}\frac{C_{\ell m}(t_j)}\ell\Phi_{\ell m}^\phi(\Omega_i),
    \label{eq:Xphi}
\end{align}
where $\Phi_{\ell m}^\theta(\Omega_i)$ and $\Phi_{\ell m}^\phi(\Omega_i)$ are the $\thetahat$- and $\phihat$-components of $\bm\Phi_{\ell m}$ as evaluated at the location $\Omega_i$ of station $i$.
The notation `$\{i|t_j\in\mathcal T_i\}$' indicates that the outer sum is over all stations $i$ which recorded a valid measurement at time $t_j$.
The inner sum is taken over%
\footnote{\label{ftnt:max_ell}%
    Because the Gauss coefficients $g_{\ell m}$ and $h_{\ell m}$ are largest for low $\ell$, higher $\ell$ terms in the sums in \eqref[s]{Xtheta} and (\ref{eq:Xphi}) become increasingly negligible.
    We choose to truncate the sums at $\ell=4$ and have explicitly verified that this choice has negligible effect on our analysis.
    In particular, the results we get by truncating at $\ell=3$ or $\ell=5$ differ negligibly from those we present with $\ell = 4$.
} %
$\ell\leq4$ and $-\ell\leq m\leq\ell$.
The coefficients $C_{\ell m}$ are computed from the Gauss coefficients $g_{\ell m}$ and $h_{\ell m}$ of the IGRF model (see \secref{igrfmodel}), per \eqref{clm}.
The IGRF model provides the values for $g_{\ell m}$ and $h_{\ell m}$ at five-year intervals from 1900 to 2020, and the values at all intermediate times are computed via linear interpolation of these coefficients.
Thus $C_{\ell m}(t_j)$ exhibits a gradual time dependence in \eqref[s]{Xtheta} and (\ref{eq:Xphi}).

Motivated by the stationarity of the noise in our time series over any given calendar year (see Appendix E\,1 in \citeR{Fedderke:2021iqw}), we take the weights $w_i^\theta(t_j)$ and $w_i^\phi(t_j)$ in \eqref[s]{Xtheta} and (\ref{eq:Xphi}) to be constant over each calendar year.
In principle, these weights could be arbitrary; however, as in \citeR{Fedderke:2021iqw} we set the weights within a calendar year in a data-driven fashion using the variances of the measured magnetic fields within each given year.
In particular,
\begin{align}
    w_i^\alpha(t)&=\lb[\frac1{N_i^a}\sum_{t_j\in\mathcal T_i^a } \lb[B_i^\alpha(t_j)\rb]^2\rb]^{-1},
\end{align}
for $\alpha=\theta,\phi$, where $\mathcal{T}_i^a$ is the subset of $\mathcal{T}_i$ contained entirely within year $a$, and $N_i^a$ is the corresponding number of data points in $\mathcal{T}_i^a$.
Note that we make use of the baseline-subtracted (zero-mean) SuperMAG field data, so that $\text{Var}\lb[B_i^{\alpha}\rb] = \langle \lb[ B_i^{\alpha} \rb]^2 \rangle$; see the detailed discussion of the appropriateness of the latter choice in Sec.~III\,C of \citeR{Fedderke:2021iqw}.
The normalizing total weights are then defined by
\begin{align}
    W^\alpha(t_j)&=\sum_{\{i|t_j\in\mathcal T_i\}}w_i^\alpha(t_j).
\end{align}

The rest of our analysis works solely with the time series $X^\theta(t_j)$ and $X^\phi(t_j)$, rather than the station-by-station data.
As mentioned at the end of \secref{derivation}, the axion dark matter has a finite coherence time $T_{\text{coh}}(m_a)$, which may be shorter than the 50-year duration of the SuperMAG dataset.
It is therefore convenient to partition our full time series $X^\theta$ and $X^\phi$ into shorter segments $X_k^\theta$ and $X_k^\phi$, each roughly the length of the coherence time $T_\text{coh}\sim2\pi/(m_av_{\textsc{dm}}^2)\sim10^6f_a^{-1}$.
Proceeding in this fashion allows us to analyze each individual segment $k$ coherently [i.e., \eqref{signal} can be assumed to be accurate over the whole duration of the segment] and then combine the individual segments' results incoherently.
As we will be interested in setting a bound at a particular axion frequency $f_a$, we take the Fourier transform of each $X_k^\theta$ and $X_k^\phi$ at $f_a$ and combine them into several two-dimensional `analysis vectors'%
\footnote{\label{ftnt:vectors}%
    Here and throughout we use $\vec x$ to denote a vector $x$ with 2 components.
} %
\begin{align}
    \vec X_k=\begin{pmatrix}
			\tilde X_k^\theta(f_a)\\[2ex]
			\tilde X_k^\phi(f_a)
		\end{pmatrix}.
	\label{eq:Xk}
\end{align}
We note that while in \citeR{Fedderke:2021iqw} it was necessary to include the Fourier transforms at $f_a\pm f_d$ [where $f_d = (\text{sidereal day})^{-1}$] in the analysis vector, this is not necessary in the present analysis, as the axion signal has no $f_d$-dependence (see the discussion in \secref{comparison}).
Therefore, the dimensionality of the analysis vector $\vec{X_k}$ in this work is significantly reduced from its 15 dimensions in the dark-photon dark-matter case considered in \citeR{Fedderke:2021iqw} to only two dimensions in the axion dark-matter case considered here.

Finally, we note that in order to efficiently make use of the Fast Fourier Transform (FFT), we must approximate $T_\text{coh}$ in such a way that many frequencies $f_a$ at which we set bounds have the same approximate coherence time.
The framework by which we choose our frequencies $f_a$ and approximate $T_\text{coh}$ in a computationally efficient manner is identical to the framework used in \citeR{Fedderke:2021iqw}.
We refer the interested reader to Sec.~V\,E of \citeR{Fedderke:2021iqw} for more details on our frequency choice and coherence-time approximation.

\subsection{Statistics of \texorpdfstring{$\vec X_k$}{the vectors Xk}}
\label{app:Xkstats}

Now that we have constructed the primary variables $\vec X_k$ for our analysis, we compute their statistics in this subsection.%
\footnote{\label{ftnt:Gaussianity}%
    As in \citeR{Fedderke:2021iqw}, we assume the variables $\vec X_k$ to be Gaussian, so that their statistics are entirely described by their expectation value and variance.
    See Appendix E\,3 of \citeR{Fedderke:2021iqw} for validation of this assumption.
} %
Let us begin with the expectation $\langle\vec X_k\rangle$ under the hypothesis of a signal with axion-photon coupling $g_{a\gamma}$.
We parametrize the axion amplitude as
\begin{align}
    c = \frac{\sqrt2\pi f_aa_0}{\sqrt{\rho_\textsc{DM}}},
    \label{eq:cdef}
\end{align}
so that $\langle|c|^2\rangle = 1$.
In the case of a true axion dark-matter signal with amplitude $a_0$ and coupling $g_{a\gamma}$, the physical field which would be measured by the SuperMAG magnetometers would be the real part of \eqref{signal}.
Because the VSH sum in \eqref{signal} is manifestly real, the measured field at $r=R$ can simply be written as%
\footnote{\label{ftnt:YPsi_modes}%
    As noted in \secref{derivation}, this expression for the magnetic field only accounts for the $\bm\Phi_{\ell m}$ contributions to the axion dark-matter signal, but $\bm Y_{\ell m}$ and $\bm\Psi_{\ell m}$ may in principle also be present.
    As in \citeR{Fedderke:2021iqw}, we ignore such additional contributions in our analysis.
    If the station locations $\Omega_i$ were uniformly distributed over the Earth's surface, and the weights were taken to be $w_i^\theta(t_j)=w_i^\phi(t_j)=1$ for all stations $i$ and times $t_j$, then \eqref[s]{Xtheta} and (\ref{eq:Xphi}) would approximate a uniform integral over the sphere and so project out any $\bm Y_{\ell m}$ and $\bm\Psi_{\ell m}$ contributions due to the VSH orthogonality relations \eqrefRange{Y_orthog}{cross_orthog}.
    Since this is not exactly the case, it is possible for other VSH modes to `leak' into our time series $X^\theta$ and $X^\phi$.
    However, this leakage would at worst affect our search at the level of $\mathcal O(1)$ factors, and so we neglect such contributions.
    See Sec.~V\,B\,1 of \citeR{Fedderke:2021iqw} for further discussion on this point.
} %
\begin{align}
    \bm B_i(t_j)&=\IM\left[ce^{-2\pi if_a t_j}\right]\sqrt{2\rho_\textsc{DM}}g_{a\gamma}R\nl
    \qquad\times \sum_{\ell,m}\frac{C_{\ell m}(t_j)}\ell\bm\Phi_{\ell m}(\Omega_i).
    \label{eq:real_signal}
\end{align}
Let us also define the time series
\begin{align}
    H^\alpha(t_j) &= \frac1{W^\alpha(t_j)}\sum_{\{i|t_j\in\mathcal T_i\}}w_i^\alpha(t_j)\left[\sum_{\ell,m}\frac{C_{\ell m}(t_j)}\ell\Phi_{\ell m}^\alpha\right]^2,
    \label{eq:Hseries}
\end{align}
for $\alpha=\theta,\phi$, and let $H^\alpha_k$ represent the subseries of these with the same sampling times as $X^\alpha_k$, and let $\tilde{H}^\alpha_k$ be their Fourier transforms.
Substituting \eqref{real_signal} into \eqref{Xk}, we find that the expectation value of our analysis vector, under the signal hypothesis, is
\begin{align}
    \langle\vec X_k\rangle&=i\sqrt{\frac{\rho_\textsc{DM}}2}g_{a\gamma}R\begin{pmatrix}
			c_k^*\tilde H^\theta_k(0)-c_k\tilde H^\theta_k(2f_a)\\[1ex]
			c_k^*\tilde H^\phi_k(0)-c_k\tilde H^\phi_k(2f_a)
		\end{pmatrix}\\[2ex]
		&\approx i\sqrt{\frac{\rho_\textsc{DM}}2}g_{a\gamma}Rc_k^*\begin{pmatrix}
			\tilde H^\theta_k(0)\\[1ex]
			\tilde H^\phi_k(0).
		\end{pmatrix}\equiv g_{a\gamma}c_k^*\vec\mu_k,
\end{align}
where $c_k$ denotes the normalized axion amplitude [\eqref{cdef}] in the $k$-th coherence time, and where at the~$\approx$~sign we have assumed that $\tilde H^\theta_k$ and $\tilde H^\phi_k$ decay rapidly with frequency, so that we may ignore higher-frequency contributions (see discussion after Eq.~(36) in \citeR{Fedderke:2021iqw}).%
\footnote{\label{ftnt:low_frequency}%
    Note that the coefficients $C_{\ell m}$ that appear in \eqref{Hseries} exhibit a time dependence.
    These coefficients thus introduce an additional time dependence which was not present in the dark-photon dark-matter case considered in \citeR{Fedderke:2021iqw}.
    As these coefficients drift by $\mathcal O(10\%)$ on the century timescale, their time dependence affects significantly lower frequencies than are relevant for our analysis.
} %

Now we turn to the variance of the $\vec X_k$ variables under the zero-signal hypothesis.
As in \citeR{Fedderke:2021iqw}, we estimate the noise in $\vec X_k$ under the assumption that it is stationary within any given calendar year.
(This assumption was validated in Appendix E\,1 of \citeR{Fedderke:2021iqw} by showing that the noise estimates from different quarters of a calendar year agreed with the full-year estimate.)
In particular, let $x^\alpha(t_j)$ denote a hypothetical realization of the noise in $X^\alpha$ (i.e., under the assumption of no signal $g_{a\gamma}=0$) over a duration $\tau$ entirely contained within calendar year $a$.
Then we define the two-sided cross-power spectral density $S^a_{\alpha\beta}$ for the year $a$ by
\begin{align}\label{eq:spectral_density}
    \langle\tilde x^\alpha(f_p)\,\tilde x^\beta(f_q)^*\rangle_{g_{a\gamma}=0} \equiv \tau S_{\alpha\beta}^a(f_p')\, \delta_{pq},
\end{align}
where $\tilde x^\alpha(f_p)$ is the Discrete Fourier Transform (DFT) of $x^{\alpha}(t_j)$ evaluated at one of the DFT frequencies $f_{p,q}$, and $\delta_{pq}$ is the Kronecker delta.
We compute the power spectral density $S^a_{\alpha\beta}$ in a data-driven manner identical to that used in Sec.~V\,C of \citeR{Fedderke:2021iqw}.
Namely, we partition the calendar year into several chunks, each of which we treat as an independent noise realization, in order to evaluate \eqref{spectral_density}.
Once the power spectral density has been calculated for each calendar year, the covariance matrix for $\vec X_k$ can be computed by combining the noise spectra from different years falling within the same coherence time:
\begin{align}
    \Sigma_k\equiv\text{Cov}(\vec X_k,\vec X_k)=\sum_aT^a_k\cdot S^a_{\alpha\beta}(f_a),
\end{align}
where $T^a_k$ is the duration of the subseries $X_k^{(m)}$ contained within calendar year $a$, so that $\sum_aT^a_k$ is the full duration of the subseries $X_k^{(m)}$, approximately given by the coherence time $T_\text{coh}$.
The quantities $\vec\mu_k$ and $\Sigma_k$ are then sufficient to describe the statistics of $\vec X_k$.

\subsection{Bayesian statistical analysis}
\label{app:bayesian}

Having computed the statistics of the $\vec X_k$ variables, we can now analyze them to set bounds on $g_{a\gamma}$.
We will do so in a Bayesian framework: we will write down a likelihood function for the $\vec X_k$; then beginning with a prior on $g_{a\gamma}$, we will utilize this likelihood function to derive a posterior on $g_{a\gamma}$.
Apart from a few $\mathcal O(1)$ numbers, the formulae in this subsection will be almost identical to those derived in Sec.~V\,D of \citeR{Fedderke:2021iqw}, but we reproduce them here for completeness.
Let us begin by writing down the likelihood function for an axion-photon coupling $g_{a\gamma}$ and normalized axion amplitudes $c_k$, in terms of the observed analysis vectors $\vec X_k$
\begin{widetext}
\begin{align}
    - \ln \LL\lb(g_{a\gamma},\{c_k\}\big|\{\vec X_k\}\rb) = \sum_k\lb(\vec X_k-g_{a\gamma}c_k^*\vec\mu_k\rb)^\dagger\Sigma_k^{-1}\lb(\vec X_k-g_{a\gamma}c_k^*\vec\mu_k\rb).
    \label{eq:Xklikelihood}
\end{align}
\end{widetext}
Here we treat each coherence time as independent, so that we may sum over the individual log-likelihoods from each coherence time.
Since $\Sigma_k$ is an Hermitian, positive-definite matrix, we may write $\Sigma_k=A_kA_k^\dagger$ for some invertible $A_k$, and define
\begin{align}
    \vec Y_k &=A_k^{-1}\vec X_k,\\
    \vec\nu_{k}&=A_k^{-1}\vec\mu_{k}.
\end{align}
In terms of these new variables, the likelihood \eqref{Xklikelihood} can then be rewritten as
\begin{align}
    -\ln \LL\lb(g_{a\gamma},\{c_k\}\big|\{\vec Y_k\}\rb)=\sum_k\lb|\vec Y_k-g_{a\gamma}c_k^*\vec\nu_k\rb|^2.
    \label{eq:Ylikelihood}
\end{align}
We can further simplify \eqref{Ylikelihood} by projecting%
\footnote{\label{ftnt:likelihood_projection}%
    It can be shown that the component orthogonal to this projection does not depend on $g_{a\gamma}$, $c_k$, or $z_k$.
    It can therefore be neglected when restricting our attention to the likelihood in terms of $z_k$.
    See the detailed discussion of this point in the same context in Sec.~V\,D\,1 of \citeR{Fedderke:2021iqw}.
} %
each term onto the direction of $\vec\nu_k$.
Namely if we make a further change of variables
\begin{align}
    s_k&=|\nu_k|,&
    z_k&=\frac{\vec\nu_k^\dagger\vec Y_k}{s_k},
    \label{eq:sz}
\end{align}
we can write the likelihood as
\begin{align}
    -\ln \LL\lb(g_{a\gamma},\{c_k\}\big|\lb\{z_k\rb\}\rb)=\sum_k\lb|z_k-g_{a\gamma}c_k^*s_k\rb|^2.
    \label{eq:Zlikelihood}
\end{align}

We are interested in setting constraints on $g_{a\gamma}$, given data in the form of $z_k$.
We are therefore not concerned with the $c_k$ amplitudes and so marginalize over them in the likelihood \eqref{Zlikelihood}.
The real and imaginary parts of $c_k$ are each Gaussian with variance 1/2 [see discussion after \eqref{cdef}].
Therefore, their likelihoods are given by
\begin{align}
    \ln \LL_k\lb(c_k\rb)=\exp\lb(-|c_k|^2\rb).
\end{align}
Using these likelihoods, we can marginalize over $c_k$ in \eqref{Zlikelihood} to find the marginalized likelihood
\begin{align}
    \LL(g_{a\gamma}\big|\{z_k\})\propto\prod_k\frac1{1+g_{a\gamma}^2s_k^2}\exp\lb(-\frac{|z_k|^2}{1+g_{a\gamma}^2s_k^2}\rb).
    \label{eq:likelihood}
\end{align}

Now that we have the likelihood for $g_{a\gamma}$ in terms of $z_k$, we can derive a posterior for $g_{a\gamma}$ given the observed values of $z_k$.
First we must begin with a prior for $g_{a\gamma}$.
As in \citeR[s]{Fedderke:2021iqw} and \cite{Centers:2019dyn}, we take the (reparametrization-invariant) objective Jeffreys prior~\cite{Cowan:2018swr}.
Formally this is defined in terms of the Fisher information matrix~\cite{Cowan:2018swr}.
In our context, it takes the form,
\begin{align}
    p(g_{a\gamma})\propto\sqrt{\sum_k\frac{4g_{a\gamma}^2s_k^4}{\lb(1+g_{a\gamma}^2s_k^2\rb)^2}}.
    \label{eq:prior}
\end{align}
Then after observing $z_k$, the posterior for $g_{a\gamma}$ becomes
\begin{align}
    p(g_{a\gamma}|\{z_k\})&\propto\LL\lb(g_{a\gamma}\big|\{z_k\}\rb)\cdot p(g_{a\gamma})\\
    &\propto\lb[\sum_k\frac{4g_{a\gamma}^2s_k^4}{\lb(1+g_{a\gamma}^2s_k^2\rb)^2}\rb]^{\frac{1}{2}} \nonumber \\
	&\quad\times 
	\prod_k\frac1{1+g_{a\gamma}^2s_k^2}\exp\lb(-\frac{|z_k|^2}{1+g_{a\gamma}^2s_k^2}\rb).
	\label{eq:posterior}
\end{align}
The normalization for \eqref{posterior} can be calculated by requiring $\int_{0}^\infty dg_{a\gamma}\, p(g_{a\gamma}|\{z_{k}\})=1$.
Finally with the appropriate normalization computed, we can set a 95\%-credible upper limit (local significance) $\hat g_{a\gamma}$ by solving
\begin{align}
    \int_0^{\hat g_{a\gamma}}dg_{a\gamma}~p(g_{a\gamma}|\{z_k\})=0.95.
\end{align}

We apply a 25\% degradation factor to our upper limit,
\begin{align}
    \hat g_{a\gamma}\rightarrow\hat g_{a\gamma}'\equiv1.25\cdot\hat g_{a\gamma}
    \label{eq:degradation}
\end{align}
to correct for the finite width of a physical axion dark-matter signal.
The reason that this degradation factor is necessary is that, thus far, we have assumed that our signal would be exactly monochromatic within a coherence time.
However, even within a coherence time, a physical signal would have finite width, comparable to the DFT frequency resolution.
This would lead to a suppression of power at the central frequency.
In other words, the limit $\hat g_{a\gamma}$ is too strong because it assumes a signal size slightly larger than what a physical signal of finite width would exhibit at its central frequency.
In \citeR{Fedderke:2021iqw}, we estimated that a 25\% degradation factor is appropriate to correct for this assumption.
Because that estimate was based on a single component of the vectorial dark-photon dark-matter signal, the same numerical estimate applies in the axion case.
Additionally, following a prescription similar to that described in Sec.~VI\,C of \citeR{Fedderke:2021iqw} (with the appropriate simplifications made for the axion case), we have confirmed explicitly by injecting a physical axion dark-matter signal that a 25\% degradation factor is sufficient for our analysis to return an upper limit on $g_{a\gamma}$ that is consistent with the injected signal parameters.
Our results in \figref{limit} show the corrected limit $\hat g_{a\gamma}'$, as defined in \eqref{degradation}.

\begin{table*}[t]
    \centering
    \begin{ruledtabular}
    \begin{tabular}{r|lll||llll|l||llll|l||l}
    No. & $f$ [mHz] &   $p_0$   &   $\sigma(p_0)$ &   $p_1$   &   $p_2$   &   $p_3$   &   $p_4$   &   $p_\text{time}$ &   $p_5$   &   $p_6$   &   $p_7$   &   $p_8$   &   $p_\text{geo}$  &   $p_\text{full}$ \\\hline
    1   &   $2.777777$   &   $6.6\times10^{-17}$   &   $6.2$   &   $0.85$   &   $0.81$   &   $0.01$   &   $0.01$   &   $6.4\times10^{-3}$   &   $1.00$   &   $0.90$   &   $0.05$   &   $0.47$   &   $0.025$   &   $1.1\times10^{-3}$   \\
    2   &   $3.333331$   &   $1.8\times10^{-12}$   &   $4.4$   &   $1.00$   &   $0.99$   &   $0.58$   &   $0.05$   &   $8.8\times10^{-4}$   &   $0.76$   &   $1.00$   &   $0.37$   &   $0.24$   &   $4.9\times10^{-4}$   &   $4.5\times10^{-6}$   \\
    3   &   $3.333335$   &   $3.0\times10^{-37}$   &   $11.5$   &   $0.99$   &   $0.91$   &   $0.30$   &   $0.02$   &   $0.013$   &   $0.43$   &   $1.00$   &   $0.23$   &   $0.10$   &   $1.1\times10^{-3}$   &   $1.3\times10^{-4}$   \\
    4   &   $3.333338$   &   $1.2\times10^{-10}$   &   $3.4$   &   $0.96$   &   $0.93$   &   $0.05$   &   $0.03$   &   $0.011$   &   $0.08$   &   $0.99$   &   $0.00$   &   $0.10$   &   $2.0\times10^{-4}$   &   $2.2\times10^{-5}$   \\
    5   &   $4.166668$   &   $3.5\times10^{-18}$   &   $6.7$   &   $0.91$   &   $0.35$   &   $0.29$   &   $0.01$   &   $0.14$   &   $1.00$   &   $0.58$   &   $0.04$   &   $0.37$   &   $0.033$   &   $0.023$   \\
    6   &   $5.000000$   &   $9.8\times10^{-62}$   &   $15.6$   &   $0.64$   &   $0.84$   &   $0.35$   &   $0.00$   &   $0.022$   &   $0.01$   &   $1.00$   &   $0.48$   &   $0.02$   &   $8.0\times10^{-6}$   &   $2.7\times10^{-6}$   \\
    7   &   $5.532406$   &   $4.5\times10^{-11}$   &   $3.6$   &   $1.00$   &   $0.08$   &   $0.07$   &   $0.32$   &   $0.016$   &   $0.46$   &   $0.54$   &   $0.89$   &   $0.20$   &   $0.74$   &   $0.091$   \\
    8   &   $5.555552$   &   $5.1\times10^{-9}$   &   $2.1$   &   $0.62$   &   $0.95$   &   $0.49$   &   $0.48$   &   $0.74$   &   $1.00$   &   $0.96$   &   $0.05$   &   $0.63$   &   $4.3\times10^{-7}$   &   $2.5\times10^{-5}$   \\
    9   &   $5.555557$   &   $6.0\times10^{-12}$   &   $4.1$   &   $0.97$   &   $0.41$   &   $0.49$   &   $0.45$   &   $0.58$   &   $1.00$   &   $0.77$   &   $0.55$   &   $1.00$   &   $2.8\times10^{-6}$   &   $6.9\times10^{-5}$   \\
    10   &   $6.666661$   &   $9.8\times10^{-17}$   &   $6.2$   &   $0.99$   &   $0.83$   &   $0.11$   &   $0.07$   &   $0.027$   &   $0.13$   &   $1.00$   &   $0.78$   &   $0.51$   &   $3.1\times10^{-7}$   &   $1.8\times10^{-7}$   \\
    11   &   $6.666668$   &   $1.7\times10^{-78}$   &   $17.9$   &   $0.19$   &   $0.96$   &   $0.34$   &   $0.04$   &   $0.12$   &   $0.63$   &   $1.00$   &   $0.31$   &   $0.08$   &   $6.3\times10^{-11}$   &   $5.5\times10^{-10}$   \\
    12   &   $6.944444$   &   $2.2\times10^{-15}$   &   $5.7$   &   $1.00$   &   $0.75$   &   $0.45$   &   $0.01$   &   $8.5\times10^{-3}$   &   $1.00$   &   $0.97$   &   $0.26$   &   $0.96$   &   $4.7\times10^{-6}$   &   $6.0\times10^{-7}$   \\
    13   &   $6.944451$   &   $1.3\times10^{-9}$   &   $2.6$   &   $0.31$   &   $0.97$   &   $0.46$   &   $0.01$   &   $0.051$   &   $1.00$   &   $0.77$   &   $0.49$   &   $0.77$   &   $0.034$   &   $9.8\times10^{-3}$   \\
    14   &   $7.777782$   &   $2.0\times10^{-9}$   &   $2.5$   &   $1.00$   &   $0.85$   &   $0.01$   &   $0.27$   &   $3.7\times10^{-3}$   &   $0.80$   &   $0.72$   &   $0.39$   &   $0.99$   &   $0.16$   &   $4.5\times10^{-3}$   \\
    15   &   $8.310182$   &   $3.4\times10^{-9}$   &   $2.3$   &   $0.81$   &   $0.14$   &   $0.02$   &   $0.76$   &   $0.14$   &   $0.81$   &   $0.25$   &   $0.12$   &   $0.98$   &   $0.11$   &   $0.068$   \\
    16   &   $8.333333$   &   $2.9\times10^{-41}$   &   $12.2$   &   $0.98$   &   $0.00$   &   $0.36$   &   $0.06$   &   $7.8\times10^{-3}$   &   $0.98$   &   $1.00$   &   $0.64$   &   $0.82$   &   $5.5\times10^{-6}$   &   $6.4\times10^{-7}$   \\
    17   &   $8.356485$   &   $3.4\times10^{-9}$   &   $2.3$   &   $0.81$   &   $0.14$   &   $0.02$   &   $0.76$   &   $0.14$   &   $0.81$   &   $0.25$   &   $0.12$   &   $0.98$   &   $0.11$   &   $0.068$   \\
    18   &   $8.888885$   &   $4.7\times10^{-12}$   &   $4.2$   &   $1.00$   &   $0.78$   &   $0.07$   &   $0.36$   &   $7.9\times10^{-3}$   &   $0.86$   &   $0.88$   &   $0.68$   &   $1.00$   &   $0.045$   &   $2.4\times10^{-3}$   \\
    19   &   $9.722221$   &   $1.6\times10^{-19}$   &   $7.1$   &   $0.95$   &   $0.96$   &   $0.29$   &   $0.00$   &   $1.5\times10^{-3}$   &   $1.00$   &   $1.00$   &   $0.36$   &   $0.99$   &   $3.7\times10^{-13}$   &   $2.5\times10^{-14}$   \\
    20   &   $10.00000$   &   $5.2\times10^{-87}$   &   $19.0$   &   $0.98$   &   $0.88$   &   $0.22$   &   $0.01$   &   $0.012$   &   $0.70$   &   $1.00$   &   $0.09$   &   $0.05$   &   $1.3\times10^{-11}$   &   $7.2\times10^{-12}$   \\
    21   &   $11.08796$   &   $7.2\times10^{-9}$   &   $2.0$   &   $1.00$   &   $0.09$   &   $0.91$   &   $0.15$   &   $2.8\times10^{-7}$   &   $0.68$   &   $0.98$   &   $1.00$   &   $0.97$   &   $1.5\times10^{-5}$   &   $7.5\times10^{-11}$   \\
    22   &   $11.11111$   &   $4.6\times10^{-16}$   &   $5.9$   &   $0.99$   &   $0.95$   &   $0.77$   &   $0.02$   &   $7.8\times10^{-3}$   &   $1.00$   &   $0.97$   &   $0.32$   &   $1.00$   &   $1.6\times10^{-13}$   &   $7.4\times10^{-14}$   \\
    23   &   $11.38889$   &   $8.5\times10^{-9}$   &   $1.9$   &   $1.00$   &   $0.81$   &   $0.46$   &   $0.09$   &   $2.6\times10^{-3}$   &   $0.99$   &   $0.30$   &   $0.82$   &   $0.96$   &   $0.042$   &   $8.6\times10^{-4}$   \\
    24   &   $11.66666$   &   $3.7\times10^{-56}$   &   $14.8$   &   $0.83$   &   $0.97$   &   $0.41$   &   $0.00$   &   $1.8\times10^{-4}$   &   $0.05$   &   $1.00$   &   $0.38$   &   $0.05$   &   $2.0\times10^{-14}$   &   $1.6\times10^{-16}$   \\
    25   &   $12.50000$   &   $3.9\times10^{-23}$   &   $8.2$   &   $1.00$   &   $0.84$   &   $0.80$   &   $0.00$   &   $4.5\times10^{-4}$   &   $1.00$   &   $1.00$   &   $0.13$   &   $0.44$   &   $5.2\times10^{-10}$   &   $6.1\times10^{-12}$   \\
    26   &   $13.33333$   &   $1.1\times10^{-49}$   &   $13.7$   &   $0.98$   &   $0.96$   &   $0.13$   &   $0.00$   &   $8.2\times10^{-7}$   &   $0.00$   &   $1.00$   &   $0.00$   &   $0.02$   &   $1.6\times10^{-14}$   &   $4.6\times10^{-19}$   \\
    27   &   $13.88889$   &   $9.0\times10^{-25}$   &   $8.6$   &   $0.75$   &   $0.97$   &   $0.02$   &   $0.00$   &   $6.9\times10^{-4}$   &   $1.00$   &   $1.00$   &   $0.00$   &   $0.08$   &   $2.2\times10^{-7}$   &   $2.7\times10^{-9}$
    \end{tabular}
    \end{ruledtabular}
    \caption{%
        Na\"ive signal candidates and associated $p$-values.
        $p_0$ indicates the local significance of the candidate in the original analysis, under the hypothesis of no signal.
        This is also translated into a global one-sided Gaussian-standard-deviation significance $\sigma(p_0)$.
        The values $p_1,\ldots,p_4$ indicate how well the signal sizes implied by four temporal subsets agree with the signal size implied by the full dataset.
        The values $p_5,\ldots,p_8$ indicate a similar agreement, but for four geographical subsets.
        Note that $p_j$ near either 0 or 1 indicates poor agreement with the full dataset.
        $p_\text{time}$ (respectively, $p_\text{geo}$) indicates the combined significance of all the temporal (geographical) checks.
        $p_\text{full}$ represents the combined significance of all eight tests.
    }
    \label{tab:resampling}
\end{table*}

\subsection{Na\"ive signal candidate re-evaluation}
\label{app:reevaluation}

In this subsection, we identify any na\"ive axion dark-matter signal candidates in our analysis and test them for robustness.
Our candidate identification and re-evaluation procedure is essentially the same as in \citeR{Fedderke:2021iqw}.
Here we briefly review the details of the procedure, as well as discuss the results of the re-evaluation.

We define a na\"ive signal candidate in a frequentist fashion as a frequency for which the observed $z_k$ are inconsistent with the absence of a signal.
According to the likelihood \eqref{likelihood} with $g_{a\gamma}=0$, each $z_k$ should have Gaussian real and imaginary parts, each with variance 1/2.
Therefore, the statistic,
\begin{align}
    Q_0=2\sum_k|z_k|^2
\end{align}
should follow a $\chi^2$-distribution.
We can compute an associated $p$-value
\begin{align}
    p_0=1-F_{\chi^2(2K_0)}[Q_0],
\end{align}
where $F_{\chi^2(\nu)}$ is the cumulative distribution function (CDF) of the $\chi^2$-distribution with $\nu$ degrees of freedom, and $K_0$ is the number of distinct segments into which we break our time series (or in other words, the number of values through which the index $k$ ranges).
We declare a frequency to be a `na\"ive signal candidate' (with 95\% global confidence) if $p_0$ is below the threshold,
\begin{align}
    p_\text{crit}=1-0.95^{1/N_f}\approx1.6\times10^{-8},
\end{align}
where $N_f\approx3.3\times10^6$ is the number of distinct frequencies in our range of interest $6\times10^{-4}\Hz<f_{A'}<\lb[ (1\text{ min})^{-1} - 6\times10^{-4}\rb]\Hz$.
Based on this criterion, we identify 27 na\"ive signal candidates; see Table I.
The global significance of each candidate can also be characterized by its equivalent one-sided Gaussian-standard-deviation
\begin{align} 
    \sigma(p_0) \equiv \sqrt{2}\, \text{erfc}^{-1}\lb[ 2\lb( 1 - \lb( 1-p_0\rb)^{N_{f}} \rb)\rb].
\end{align}

We however do not immediately consider these 27 na\"ive signal candidates to necessarily be promising axion dark-matter signals: we first
re-evaluate each na\"ive signal candidate by performing several robustness checks to test their temporal consistency and spatial uniformity.
Any axion dark-matter signal should be present over the entire 50-year duration of the SuperMAG dataset and should be present in all 508 stations.
We therefore re-perform our analysis on eight subsets of the full SuperMAG dataset: four temporal subsets and four geographical subsets.
The four temporal subsets consist only of the data from the ranges of years 1970--1982, 1983--1994, 1995--2007, and 2008--2019, respectively.
Meanwhile, the geographical subsets consist of the data from four random disjoint subsets of stations.%
\footnote{\label{ftnt:geographic}%
    For the first six years of data, at least one of the geographical subsets do not contain an active station.
    Therefore, for the geographic resampling tests, we discard the first six years of data and only use the years 1986--2019.
} %
For each frequency and each subset, the robustness test consists of checking whether the signal appears in a way that is consistent with the posterior on $g_{a\gamma}$ derived from the original analysis.

More specifically, let $s_{k,j}$ and $z_{k,j}$ denote the quantities defined in \eqref{sz}, but computed using subset $j$ of the full SuperMAG dataset (where $j=1,\ldots,4$ refers to the temporal subsets and $j=5,\ldots,8$ refers to the geographical subsets).
From the likelihood \eqref{likelihood}, we can see that the statistic,
\begin{align}
    Q_j(g_{a\gamma})=\sum_k\frac{2|z_{k,j}|^2}{1+g_{a\gamma}^2s_{k,j}^2}
\end{align}
should follow a $\chi^2$-distribution under the assumption of a signal with coupling $g_{a\gamma}$.
We can again compute an associated $p$-value
\begin{align}
    p_j(g_{a\gamma})=F_{\chi^2(2K_j)}[Q_j(g_{a\gamma})],
\end{align}
where $K_j$ is the number of time-series segments in the $j^\text{th}$ resampling analysis (which may differ from $K_0$ due to the reduced total duration of the subsets).
To test for consistency with the original analysis, we then weight the $p$-values for each resampling test according to the posterior $p(g_{a\gamma}|\{z_k\})$ derived in the original analysis
\begin{align}
    p_j=\int dg_{a\gamma}~p(g_{a\gamma}|\{z_k\})\cdot p_j(g_{a\gamma}).
\end{align}
Note that $p_j$ near either 0 or 1 indicates disagreement with the original analysis, as they imply that the inferred signal present in the subset is much smaller or larger, respectively, than the signal in the full dataset.
Finally, we can combine the $p$-values from all $n=8$ tests into a single statistic using a two-tailed version of Fisher's method~\cite{Fisher:1958iqe,10.2307/2681650,10.2307/2529826,KOST2002183}.
Specifically we define the combined $\chi^2$-statistic 
\begin{align}
    Q_\text{full}=-2\sum_j\ln\left(2\cdot\text{min}\{p_j,1-p_j\}\right),
    \label{eq:Qfull}
\end{align}
and its associated $p$-value
\begin{align}
    p_\text{full}=1-F_{\chi^2(2n)}(Q_\text{full}) \qquad [n=8].
    \label{eq:pfull}
\end{align}
In addition to $p_\text{full}$, we also compute combined $p$-values $p_\text{time}$ and $p_\text{geo}$ for the temporal-only and geographical-only tests, respectively, by restricting the sum in \eqref{Qfull} to the appropriate tests and setting $n=4$ in \eqref{pfull}.

As noted and discussed at greater length in Sec.~VI\,B of \citeR{Fedderke:2021iqw}, the temporal and geographical tests here are not entirely independent, as the temporal and geographical subsets generally contain overlapping data.
Thus our na\"ive procedure of combining the results of all the tests as if they were independent in \eqref[s]{Qfull} and (\ref{eq:pfull}) is not exactly accurate.
Therefore, rather than immediately rejecting (at 95\% confidence) all candidates with $p_\text{full}<0.05$, we instead only reject candidates with $p_\text{full}<0.01$, and deem candidates with $0.01<p_\text{full}<0.05$ to be in `strong tension' with our resampling checks.

\tabref{resampling} shows the results of our resampling analysis for all 27 na\"ive signal candidates.
Of the 27 candidates, 23 have $p_\text{full}<0.01$, and so we reject them as axion dark-matter signals on the basis of our combined resampling checks.
Although none of the other four candidates constitute strong evidence for axion dark matter, they require further discussion.
Here we review each one in detail.

\paragraphdash{Candidate 5}
This candidate has $\sigma(p_0) = 6.7$, but $0.01<p_\text{full}<0.05$.
While this signal is thus large, it exhibits strong (but not necessarily definitive) tension with the combined spatio-temporal robustness check.
In addition, this candidate is also in strong tension with the geographical tests alone ($p_\text{geo}=0.033<0.05$). 
We also note in passing that this candidate appears at the DFT frequency closest to half of the Nyquist frequency.
Based on the tensions with the robustness tests, we do not consider this a robust axion dark-matter candidate.

\paragraphdash{Candidate 7}
This candidate has $\sigma(p_0)=3.6$ and exhibits good agreement with both the geographical tests ($p_\text{geo}=0.74$) and the combined tests ($p_\text{full}=0.091$).
It however has $0.01<p_\text{time}<0.05$, so that it is in strong tension with the temporal tests.
We therefore consider this candidate to be in strong tension with some of our robustness tests, but we cannot definitively rule it out.

\paragraphdash{Candidates 15 and 17}
These two candidates are reflections of each other across the Nyquist frequency (and thus should be thought of as a single candidate).
They exhibit good agreement with all the robustness tests ($p_\text{time}=0.14$, $p_\text{geo}=0.11$, $p_\text{full}=0.068$).
However, they exhibit a fairly weak global significance of $\sigma(p_0)=2.3$, and so we do not consider these to be strong axion dark-matter candidates.
These candidates would be interesting to revisit in an analysis of the one-second SuperMAG dataset to see if they increase in significance.
They also motivate the development of further checks in future work to verify the physicality of our na\"ive signal candidates.

In summary, we find that none of the 27 na\"ive signal candidates constitute both strong and robust evidence for an axion dark-matter signal.
However, four of the candidates warrant further investigation in follow-up work, such as an analysis of the one-second resolution SuperMAG dataset.

\bibliographystyle{JHEP}
\bibliography{references.bib}

\end{document}